# Soft-Input Soft-Output Single Tree-Search Sphere Decoding

Christoph Studer and Helmut Bölcskei

### Abstract

Soft-input soft-output (SISO) detection algorithms form the basis for iterative decoding. The computational complexity of SISO detection often poses significant challenges for practical receiver implementations, in particular in the context of multiple-input multiple-output (MIMO) wireless communication systems. In this paper, we present a low-complexity SISO sphere-decoding algorithm, based on the single tree-search paradigm proposed originally for soft-output MIMO detection in Studer, *et al., IEEE J-SAC, 2008*. The new algorithm incorporates clipping of the extrinsic log-likelihood ratios (LLRs) into the tree-search, which results in significant complexity savings and allows to cover a large performance/complexity tradeoff region by adjusting a single parameter. Furthermore, we propose a new method for correcting approximate LLRs —resulting from sub-optimal detectors— which (often significantly) improves detection performance at low additional computational complexity.

This paper was presented in part at the IEEE International Symposium on Information Theory, Toronto, Ontario, Canada, July 2008. This work was partially supported by the STREP project No. IST-026905 (MASCOT) within the Sixth Framework Programme (FP6) of the European Commission and by the Swiss Innovation Promotion Agency (KTI/CTI) project No. 9268.1 PFNM-NM.

The authors are with the Communication Technology Laboratory, ETH Zurich, CH-8092 Zurich, Switzerland (e-mail: {studerc,boelcskei}@nari.ee.ethz.ch).

## I. INTRODUCTION

Soft-input soft-output (SISO) detection constitutes the basis for iterative decoding in multiple-input multiple-output (MIMO) systems, which, in general, achieves significantly better (error-rate) performance than decoding based on hard-output or soft-output-only detection algorithms [1]. Unfortunately, this performance gain comes at the cost of a significant (often prohibitive in terms of practical implementation) increase in computational complexity.

Various SISO detection algorithms for MIMO systems offering different performance/complexity tradeoffs have been proposed in the literature, see e.g., [1]–[6]. However, implementing different algorithms, each optimized for a maximum allowed detection effort or for a particular system configuration, would entail considerable circuit complexity. A practical SISO detector for MIMO systems should therefore cover a wide range of performance/complexity tradeoffs and be easily adjustable through a *single* tunable detection algorithm.

Soft-output single tree-search (STS) sphere decoding (SD) in combination with log-likelihood ratio (LLR) clipping [7] has been demonstrated to be suitable for VLSI implementation and allows to conveniently tune detection performance between maximum-likelihood (ML) a posteriori probability (APP) soft-output detection and (low-complexity) hard-output detection. The STS-SD concept is therefore a promising basis for efficient SISO detection in MIMO systems.

*Contributions:* We describe a SISO STS-SD algorithm that is tunable between max-log optimal SISO and hard-output maximum a posteriori (MAP) detection performance. To this end, we extend the soft-output STS-SD algorithm introduced in [7], [8] by a max-log optimal a priori information processing method, which significantly reduces the tree-search complexity compared to, e.g., [3], [5], [6], [9], [10], and avoids the computation of transcendental functions. The basic idea for complexity reduction and to achieve tunability of the algorithm is to incorporate clipping of the extrinsic LLRs into the tree search. This requires that the list administration concept and the tree-pruning criterion proposed for soft-output STS-SD in [7] be suitably modified. We furthermore propose a method for compensation of self-interference in the LLRs—caused by channel-matrix regularization—directly in the tree search. In addition, we describe a new method for correcting approximate LLRs—resulting from sub-optimal detectors—which (often significantly) improves detection performance at low additional computational complexity. Simulation results show that the resulting SISO STS-SD algorithm operates close to outage capacity at remarkably low computational complexity. In addition, the algorithm offers a significantly larger performance/complexity tradeoff region than the soft-output STS-SD algorithm proposed in [7].

*Notation:* Matrices are set in boldface capital letters, vectors in boldface lowercase letters. The superscripts $^T$ and $^H$ stand for transpose and conjugate transpose, respectively. We write $A_{i,j}$ for the entry in the $i$th row and $j$th column of the matrix $\mathbf{A}$ and $b_i$ for the $i$th entry of the vector $\mathbf{b} = [\, b_1 \; \cdots \; b_N \,]^T$. The $\ell^2$-norm of the vector $\mathbf{b}$ is denoted by $\|\mathbf{b}\|$. $\mathbf{I}_N$ and $\mathbf{0}_{M \times N}$ refer to the $N \times N$ identity matrix and the $M \times N$ all-zero matrix, respectively. Slightly abusing common terminology, we call an $M \times N$ matrix $\mathbf{A}$, where $M \geq N$, satisfying $\mathbf{A}^H \mathbf{A} = \mathbf{I}_N$, unitary. $|\mathcal{O}|$ denotes the cardinality of the set $\mathcal{O}$. The probability of an event $\mathcal{Z}$ is referred to as $\mathrm{P}[\mathcal{Z}]$, the probability density function of a continuous random variable (RV) $z$ is denoted by $\mathrm{p}(z)$ and $\mathbb{E}[Z]$ stands for the expectation of the RV $Z$. $\overline{x}$ is the binary complement of $x \in \{+1, -1\}$, i.e., $\overline{x} = -x$.

*Outline:* The remainder of this paper is organized as follows. Section II reviews the transformation of soft-input soft-output MIMO detection into a tree-search problem and presents new methods for tightening of the tree-pruning criterion and for incorporating a priori information into the tree search. Section III describes the new SISO STS-SD algorithm. In Section IV, we propose a method for compensating the impact of channel-matrix regularization on LLRs directly in the tree search. A new technique for computationally efficient correction of approximate LLRs —resulting from the max-log approximation, channel-matrix regularization, and early termination [7], [11]— is presented in Section V. Simulation results are provided in Section VI. We conclude in Section VII.

## II. Soft-Input Soft-Output Sphere Decoding

Consider a MIMO system with $M_{\mathrm{T}}$ transmit and $M_{\mathrm{R}} \geq M_{\mathrm{T}}$ receive antennas. The coded bit-stream to be transmitted is mapped to (a sequence of) $M_{\mathrm{T}}$-dimensional transmit symbol vectors $\mathbf{s} \in \mathcal{O}^{M_{\mathrm{T}}}$, where $\mathcal{O}$ stands for the underlying complex scalar constellation[1] and $|\mathcal{O}| = 2^Q$. Each symbol vector $\mathbf{s}$ is associated with a label vector $\mathbf{x}$ containing $M_{\mathrm{T}}Q$ binary values chosen from the set $\{+1, -1\}$ where the null element (0 in binary logic) of GF(2) corresponds to $+1$.

---

[1]The algorithm developed in this paper can also be formulated for the case where different constellations are used on different transmit antennas. However, for the sake of simplicity of exposition, we restrict ourselves to employing the same constellation on all transmit antennas.

The corresponding bits are denoted by $x_{i,b}$, where the indices $i$ and $b$ refer to the $b$th bit in the binary label of the $i$th entry of the symbol vector $\mathbf{s} = [\, s_1 \ \cdots \ s_{M_T} \,]^T$. The associated complex baseband input-output relation is given by

$$\mathbf{y} = \mathbf{Hs} + \mathbf{n} \tag{1}$$

where $\mathbf{H}$ stands for the $M_R \times M_T$ channel matrix, $\mathbf{y}$ is the $M_R$-dimensional received signal vector, and $\mathbf{n}$ is an i.i.d. circularly symmetric complex Gaussian distributed $M_R$-dimensional noise vector with variance $N_o$ per complex entry. Different transmit powers on the individual transmit antennas are assumed to be absorbed in the channel matrix $\mathbf{H}$, which—including the corresponding scaling factors—will be referred to as the physical MIMO channel. Throughout the paper, we consider coherent detection, i.e., the receiver knows the realization of the channel matrix $\mathbf{H}$ perfectly.

### A. Max-Log LLR Computation as a Tree Search

Coherent SISO detection for MIMO systems requires computation of the LLRs [1]

$$L_{i,b} \triangleq \log\left( \frac{\mathrm{P}[x_{i,b} = +1 \,|\, \mathbf{y}, \mathbf{H}]}{\mathrm{P}[x_{i,b} = -1 \,|\, \mathbf{y}, \mathbf{H}]} \right) \tag{2}$$

for all bits $i = 1, \ldots, M_T$, $b = 1, \ldots, Q$, in the label $\mathbf{x}$. Bayes's theorem applied to (2) leads to the equivalent formulation

$$
\begin{aligned}
L_{i,b} = \log &\left( \sum_{\mathbf{s} \in \mathcal{X}_{i,b}^{(+1)}} \mathrm{p}(\mathbf{y} \,|\, \mathbf{s}, \mathbf{H}) \, \mathrm{P}[\mathbf{s}] \right) \\
- \log &\left( \sum_{\mathbf{s} \in \mathcal{X}_{i,b}^{(-1)}} \mathrm{p}(\mathbf{y} \,|\, \mathbf{s}, \mathbf{H}) \, \mathrm{P}[\mathbf{s}] \right)
\end{aligned}
\tag{3}
$$

where $\mathcal{X}_{i,b}^{(+1)}$ and $\mathcal{X}_{i,b}^{(-1)}$ are the sets of symbol vectors that have the bit corresponding to the indices $i$ and $b$ equal to $+1$ and $-1$, respectively, $\mathrm{P}[\mathbf{s}]$ corresponds to the prior, and

$$\mathrm{p}(\mathbf{y} \,|\, \mathbf{s}, \mathbf{H}) = \frac{1}{(\pi N_o)^{M_R}} \exp\left( -\frac{\|\mathbf{y} - \mathbf{Hs}\|^2}{N_o} \right).$$

Straightforward evaluation of (3) requires the computation of $|\mathcal{O}|^{M_T}$ Euclidean distances per LLR value, which, in general, leads to prohibitive computational complexity. We therefore employ the standard max-log approximation[2] on (3), which enables us to reformulate the LLR computation problem as a weighted tree-search problem that can be solved efficiently using the SD algorithm [7], [8], [13]–[20]. To this end, the channel matrix $\mathbf{H}$ is first QR-decomposed according to $\mathbf{H} = \mathbf{QR}$, where the $M_\mathrm{R} \times M_\mathrm{T}$ matrix $\mathbf{Q}$ is unitary and the $M_\mathrm{T} \times M_\mathrm{T}$ upper-triangular matrix $\mathbf{R}$ has real-valued positive entries on its main diagonal. Left-multiplying (1) by $\mathbf{Q}^H$ leads to the modified input-output relation

$$\tilde{\mathbf{y}} = \mathbf{Rs} + \mathbf{Q}^H \mathbf{n} \tag{4}$$

where $\tilde{\mathbf{y}} = \mathbf{Q}^H \mathbf{y}$ and $\mathbf{Q}^H \mathbf{n}$ is also i.i.d. circularly symmetric complex Gaussian with variance $N_\mathrm{o}$ per complex entry. In the following, we consider an iterative MIMO decoder as depicted in Fig. 1. The soft-input soft-output MIMO detector computes *intrinsic* max-log LLRs according to [1]

$$
\begin{aligned}
L_{i,b}^{\mathrm{D}} \triangleq \min_{\mathbf{s} \in \mathcal{X}_{i,b}^{(-1)}} & \left\{ \frac{1}{N_\mathrm{o}} \|\tilde{\mathbf{y}} - \mathbf{Rs}\|^2 - \log \mathrm{P}[\mathbf{s}] \right\} \\
& - \min_{\mathbf{s} \in \mathcal{X}_{i,b}^{(+1)}} \left\{ \frac{1}{N_\mathrm{o}} \|\tilde{\mathbf{y}} - \mathbf{Rs}\|^2 - \log \mathrm{P}[\mathbf{s}] \right\},
\end{aligned}
\tag{5}
$$

where the prior $\mathrm{P}[\mathbf{s}]$ is, e.g., delivered by an outer channel decoder in the form of *a priori* LLRs

$$L_{i,b}^{\mathrm{A}} \triangleq \log \left( \frac{\mathrm{P}[x_{i,b} = +1]}{\mathrm{P}[x_{i,b} = -1]} \right), \quad \forall i, b.$$

Based on the intrinsic LLRs in (5), the detector computes the *extrinsic* LLRs

$$L_{i,b}^{\mathrm{E}} \triangleq L_{i,b}^{\mathrm{D}} - L_{i,b}^{\mathrm{A}}, \quad \forall i, b, \tag{6}$$

that are passed to a subsequent SISO channel decoder. Note that we neglected the additive constant in each of the two minima in (5) that results from the part of the noise $\mathbf{n}$ that is orthogonal to the range space of $\mathbf{H}$. This is possible as the constant in question is independent of $\mathbf{s}$ and, hence, cancels out upon taking the difference in (5).

For each bit, one of the two minima in (5) corresponds to

$$\lambda^{\mathrm{MAP}} \triangleq \frac{1}{N_\mathrm{o}} \left\| \tilde{\mathbf{y}} - \mathbf{Rs}^{\mathrm{MAP}} \right\|^2 - \log \mathrm{P}\left[ \mathbf{s}^{\mathrm{MAP}} \right] \tag{7}$$

---

[2]The max-log approximation corresponds to $\log \left( \sum_k \exp(a_k) \right) \approx \max_k \{a_k\}$ and entails a performance loss compared to using the exact LLRs in (3). As shown in [12], this loss is small, in general.

which is associated with the MAP solution of the MIMO detection problem

$$\mathbf{s}^{\mathrm{MAP}} = \arg\min_{\mathbf{s} \in \mathcal{O}^{M_{\mathrm{T}}}} \left\{ \frac{1}{N_{\mathrm{o}}} \big\| \tilde{\mathbf{y}} - \mathbf{R}\mathbf{s} \big\|^2 - \log \mathrm{P}[\mathbf{s}] \right\}. \tag{8}$$

The other minimum in (5) can be computed as

$$\lambda_{i,b}^{\overline{\mathrm{MAP}}} \triangleq \min_{\mathbf{s} \in \mathcal{X}_{i,b}^{\left(\overline{x_{i,b}^{\mathrm{MAP}}}\right)}} \left\{ \frac{1}{N_{\mathrm{o}}} \big\| \tilde{\mathbf{y}} - \mathbf{R}\mathbf{s} \big\|^2 - \log \mathrm{P}[\mathbf{s}] \right\} \tag{9}$$

where $\overline{x_{i,b}^{\mathrm{MAP}}}$ denotes the (bit-wise) *counter-hypothesis* to the MAP hypothesis. With the definitions (7) and (9), the intrinsic max-log LLRs in (5) can be written ($\forall i, b$) in compact form as

$$L_{i,b}^{\mathrm{D}} = \begin{cases} \lambda_{i,b}^{\overline{\mathrm{MAP}}} - \lambda^{\mathrm{MAP}}, & x_{i,b}^{\mathrm{MAP}} = +1 \\ \lambda^{\mathrm{MAP}} - \lambda_{i,b}^{\overline{\mathrm{MAP}}}, & x_{i,b}^{\mathrm{MAP}} = -1. \end{cases} \tag{10}$$

We can therefore conclude that efficient max-log-optimal soft-input soft-output MIMO detection reduces to efficiently identifying $\mathbf{s}^{\mathrm{MAP}}$, $\lambda^{\mathrm{MAP}}$, and $\lambda_{i,b}^{\overline{\mathrm{MAP}}}$ ($\forall i, b$).

We next define the partial symbol vectors (PSVs) $\mathbf{s}^{(i)} = [\, s_i \; \cdots \; s_{M_{\mathrm{T}}} \,]^T$ and note that they can be arranged in a tree that has its root just above level $i = M_{\mathrm{T}}$ and leaves, on level $i = 1$, which correspond to symbol vectors $\mathbf{s}$. The binary-valued label vector associated with $\mathbf{s}^{(i)}$ will be denoted by $\mathbf{x}^{(i)}$. The distances

$$d(\mathbf{s}) = \frac{1}{N_{\mathrm{o}}} \big\| \tilde{\mathbf{y}} - \mathbf{R}\mathbf{s} \big\|^2 - \log \mathrm{P}[\mathbf{s}] \tag{11}$$

in (7) and (9) can be computed recursively if the following factorization holds:

$$\mathrm{P}[\mathbf{s}] = \prod_{i=1}^{M_{\mathrm{T}}} \mathrm{P}\big[ \mathbf{s}^{(i)} \big], \tag{12}$$

which is assumed from now on. Note that in practice, the symbols $s_i$ ($i = 1, \ldots, M_{\mathrm{T}}$) are often statistically independent across spatial streams; this satisfies (12) trivially with $\mathrm{P}[\mathbf{s}] = \prod_{i=1}^{M_{\mathrm{T}}} \mathrm{P}[s_i]$. We can now rewrite (11) as

$$d(\mathbf{s}) = \sum_{i=1}^{M_{\mathrm{T}}} \left( \frac{1}{N_{\mathrm{o}}} \bigg| \tilde{y}_i - \sum_{j=i}^{M_{\mathrm{T}}} R_{i,j} s_j \bigg|^2 - \log \mathrm{P}\big[ \mathbf{s}^{(i)} \big] \right)$$

which can be evaluated recursively as $d(\mathbf{s}) = d_1$, with the partial distances (PDs)

$$d_i = d_{i+1} + |e_i|, \quad i = M_{\mathrm{T}}, \ldots, 1,$$

the initialization $d_{M_{\mathrm{T}}+1} = 0$, and the distance increments (DIs)

$$|e_i| = \frac{1}{N_{\mathrm{o}}} \left| \tilde{y}_i - \sum_{j=i}^{M_{\mathrm{T}}} R_{i,j} s_j \right|^2 - \log \mathrm{P}\left[\mathbf{s}^{(i)}\right]. \tag{13}$$

Note that the DIs are non-negative since the prior terms satisfy $-\log \mathrm{P}\left[\mathbf{s}^{(i)}\right] \geq 0$. The dependence of the PD $d_i$ on the symbol vector $\mathbf{s}$ is, thanks to the upper triangularity of $\mathbf{R}$ and the assumption (12), only through the PSV $\mathbf{s}^{(i)}$. Thus, the MAP detection problem and the computation of the intrinsic max-log LLRs has been transformed into a tree-search problem: PSVs and PDs are associated with nodes, branches correspond to DIs. For brevity, we shall often say "the node $\mathbf{s}^{(i)}$" to refer to the node corresponding to the PSV $\mathbf{s}^{(i)}$. We shall furthermore use $d\left(\mathbf{s}^{(i)}\right)$ and $d\left(\mathbf{x}^{(i)}\right)$ interchangeably to denote $d_i$. Each path from the root node down to a leaf node corresponds to a symbol vector $\mathbf{s} \in \mathcal{O}^{M_{\mathrm{T}}}$. The result in (7) and (9) corresponds to the leaf associated with the smallest metric in $\mathcal{O}^{M_{\mathrm{T}}}$ and $\mathcal{X}_{i,b}^{\left(\overline{x_{i,b}^{\mathrm{MAP}}}\right)}$, respectively. The SISO STS-SD algorithm uses elements of Schnorr-Euchner SD (SESD) [15], [21], briefly summarized as follows: The search in the weighted tree is constrained to nodes which lie within a radius[3] $r$ around $\tilde{\mathbf{y}}$ and tree traversal is performed depth-first, visiting the children of a given node in ascending order of their PDs. A node $\mathbf{s}^{(i)}$ with PD $d_i$ can be pruned (along with the entire subtree originating from this node) whenever the tree-pruning criterion

$$d_i \geq r^2 \tag{14}$$

is satisfied. In the remainder of this paper, (14) is referred to as the "standard pruning criterion."

The radius $r$ has to be chosen sufficiently large such that the SD algorithm finds at least the MAP solution. Choosing $r$ too large, leads to high complexity as a large number of nodes do not satisfy the pruning criterion. In order to avoid the problem of choosing a suitable radius $r$ altogether, we employ a technique known as radius reduction [21], which consists of initializing the algorithm with $r = \infty$, and performing the update $r^2 \leftarrow d(\mathbf{s})$ whenever a valid leaf node $\mathbf{s}$ has been found.

The complexity measure used in the remainder of the paper corresponds to the total number of nodes visited by the detector, including the leaf nodes, but excluding the root node. Note that this measure was shown in [22] to be representative of the hardware complexity of a VLSI implementation of hard-output SESD.

---

[3]Note that $r$ corresponds to the radius of a hypersphere if the prior satisfies $\mathrm{P}[\mathbf{s}] = 0$.

## B. Tightening of the Tree-Pruning Criterion

Tightening of the tree-pruning criterion (14), i.e., reduction of the right-hand side (RHS) of (14), without sacrificing (max-log) optimality is highly desirable as it reduces the (tree-search) complexity. Such a reduction can be accomplished, for example, through techniques based on semi-definite relaxation and $H^\infty$-estimation theory as proposed in [23]. Unfortunately, these approaches entail, in general, a high computational complexity and are, hence, not well-suited for practical (VLSI) implementation.

In the following, we propose an alternative approach which relies on the observation that the DIs (13) contain a —generally non-zero— bias given by

$$|b_i| \triangleq \min_{\mathbf{s}^{(i)} \in \mathcal{O}^{M_T+1-i}} |e_i|, \quad i = 1, \dots, M_T. \tag{15}$$

Consider the case where the detector stands at node $\mathbf{s}^{(i)}$ on level $i$ with corresponding PD $d_i$. All leaf-level PDs $d_1$ that can be reached from the node $\mathbf{s}^{(i)}$ satisfy

$$d_1 \geq d_i + \sum_{j=1}^{i-1} |b_j|. \tag{16}$$

At level $i$, we can therefore prune every node that satisfies a tightened version of the tree-pruning criterion in (14), namely

$$d_i \geq r^2 - \sum_{j=1}^{i-1} |b_j|. \tag{17}$$

Computation of the bias term (15) requires enumeration of $|e_i|$ over all $\mathbf{s}^{(i)} \in \mathcal{O}^{M_T+1-i}$, which, in general, leads to prohibitive computational complexity. The major portion of this complexity is caused by the computation of the Euclidean distance-term $\frac{1}{N_o}|\tilde{y}_i - \sum_{j=i}^{M_T} R_{i,j} s_j|^2$ in (13), whose contribution to the bias (15), as it turns out (corresponding simulation results are shown in Section VI-A1), is negligible. Hence, we only consider the contribution to $|b_i|$ caused by the prior term $-\log \mathrm{P}[\mathbf{s}^{(i)}]$ and we define accordingly

$$|p_i| = \min_{\mathbf{s}^{(i)} \in \mathcal{O}^{M_T+1-i}} \left\{ -\log \mathrm{P}[\mathbf{s}^{(i)}] \right\}. \tag{18}$$

The corresponding tightened tree-pruning criterion is then given by

$$d_i \geq r^2 - \sum_{j=1}^{i-1} |p_j|. \tag{19}$$

For the case of the individual symbols $s_i$ ($i = 1, \ldots, M_T$) being statistically independent, i.e., $P[\mathbf{s}] = \prod_{i=1}^{M_T} P[s_i]$, we have $|p_i| = \min_{s_i \in \mathcal{O}} \{ -\log P[s_i] \}$, so that computation of the RHS of (18) results in significantly smaller complexity than that required to compute the RHS of (15).

We emphasize that using the tightened tree-pruning criterion (19) preserves max-log optimality and leads, in general, to significant complexity savings, when compared to the standard pruning criterion (14), which is widely adopted in the literature [3], [5], [6], [9], [10]. To see this, consider the case where all constellation points are equally likely[4], i.e., $P[s_i] = |\mathcal{O}|^{-1}$ for all $s_i \in \mathcal{O}$ and $i = 1, \ldots, M_T$. The corresponding total bias from level $i$ down to the leaf level is given by $\sum_{j=1}^{i-1} |p_j| = (i-1) \log |\mathcal{O}|$, which can be large, especially for nodes close to the root. Since pruning at and close to the root level, has, in general, significant impact on the number of nodes visited in the tree search, the tightened tree-pruning criterion (19) can lead to a major complexity reduction. Corresponding simulation results are provided in Section VI-A2.

### C. Tree Search in the Case of Statistically Independent Bits

We have seen above that statistical independence among individual symbols enables us to tighten the tree-pruning criterion at low additional computational complexity. For bit-interleaved coded modulation [24], in addition the bits $x_{i,b}$ ($i = 1, \ldots, M_T$, $b = 1, \ldots, Q$) are statistically independent. As shown next, this independence on the bit-level can be exploited to get further reductions in computational complexity. To see this, consider the case where the MIMO detector obtains a priori LLRs $L_{i,b}^A$ ($\forall i, b$) from an external device, e.g., a SISO channel decoder as depicted in Fig. 1. We then have [25]

$$P[s_i] = \prod_{b:x_{i,b}=+1} \frac{\exp\left(L_{i,b}^A\right)}{1 + \exp\left(L_{i,b}^A\right)} \prod_{b:x_{i,b}=-1} \frac{1}{1 + \exp\left(L_{i,b}^A\right)}$$

which can be reformulated in more compact form as

$$P[s_i] = \prod_{b=1}^{Q} \frac{\exp\left(\frac{1}{2}\left(1 + x_{i,b}\right)L_{i,b}^A\right)}{1 + \exp\left(L_{i,b}^A\right)}. \tag{20}$$

The contribution of the a priori LLRs to the prior term in the DIs in (13) can then be obtained from (20) as

$$-\log P[s_i] = \tilde{K}_i - \sum_{b=1}^{Q} \frac{1}{2} x_{i,b} L_{i,b}^A \tag{21}$$

---

[4]This, for example, is the case when no a priori information is available and all transmitted bits are equally likely.

where the constants

$$\tilde{K}_i = \sum_{b=1}^{Q} \left( \frac{1}{2} \left| L_{i,b}^{\mathrm{A}} \right| + \log \left( 1 + \exp \left( - \left| L_{i,b}^{\mathrm{A}} \right| \right) \right) \right) \tag{22}$$

are independent of the binary-valued variables $x_{i,b}$ and $\tilde{K}_i > 0$ for $i = 1, \ldots, M_{\mathrm{T}}$. Because of $-\log \mathrm{P}[s_i] \geq 0$, we can trivially infer from (21) that $\tilde{K}_i - \sum_{b=1}^{Q} \frac{1}{2} x_{i,b} L_{i,b}^{\mathrm{A}} \geq 0$. From (10) it follows that constant terms (i.e., terms that are independent of the variables $x_{i,b}$ and hence of s) in (7) and (9) cancel out in the computation of the intrinsic LLRs $L_{i,b}^{\mathrm{D}}$ ($\forall i, b$) and can therefore be neglected. A straightforward method to avoid the hardware-inefficient task of computing transcendental functions in (22) is to set $\tilde{K}_i = 0$ in the computation of (21). This can, however, lead to branch metrics that are not necessarily non-negative, which would inhibit pruning of the search tree. On the other hand, modifying the DIs in (13) by setting

$$|e_i| \triangleq \frac{1}{N_{\mathrm{o}}} \left| \tilde{y}_i - \sum_{j=i}^{M_{\mathrm{T}}} R_{i,j} s_j \right|^2 + K_i - \sum_{b=1}^{Q} \frac{1}{2} x_{i,b} L_{i,b}^{\mathrm{A}} \tag{23}$$

with $K_i = \sum_{b=1}^{Q} \frac{1}{2} \left| L_{i,b}^{\mathrm{A}} \right|$ also avoids computing transcendental functions while guaranteeing that, thanks to $\left| L_{i,b}^{\mathrm{A}} \right| - x_{i,b} L_{i,b}^{\mathrm{A}} \geq 0$ ($\forall i, b$), the so obtained branch metrics are non-negative. Furthermore, as $\tilde{K}_i \geq K_i$, using the modified DIs (often significantly) reduces the (tree-search) complexity compared to that implied by (13) using (21) and, thanks to (10), still yields max-log-optimal LLRs. The reason for complexity reduction when using the modified DIs (23) lies in the modified prior term being bias-free, i.e.,

$$\min_{s_i \in \mathcal{O}} \left\{ K_i - \sum_{b=1}^{Q} \frac{1}{2} x_{i,b} L_{i,b}^{\mathrm{A}} \right\} = 0, \quad \forall i, \tag{24}$$

which directly leads to tight tree pruning using the standard pruning criterion in (14) and hence, avoids explicit evaluation of (18).

Note that in [5, Eq. 9], the prior term (21) was approximated as

$$-\log \mathrm{P}[s_i] \approx \sum_{b=1}^{Q} \frac{1}{2} \left( \left| L_{i,b}^{\mathrm{A}} \right| - x_{i,b} L_{i,b}^{\mathrm{A}} \right)$$

for $\left| L_{i,b}^{\mathrm{A}} \right| > 2$ ($b = 1, \ldots, Q$) which corresponds exactly to what was done here in order to arrive at (23). It is important, though, to realize that using the modified DIs (23) does *not* lead to an approximation of (10), as only differences are considered in the intrinsic max-log LLR computation and the neglected $\log(\cdot)$-term does not depend on $x_{i,b}$.

## III. EXTRINSIC LLR COMPUTATION IN A SINGLE TREE SEARCH

Computing the intrinsic max-log LLRs in (10) requires to determine $\lambda^{\mathrm{MAP}}$ and the metrics $\lambda_{i,b}^{\overline{\mathrm{MAP}}}$ associated with the counter-hypotheses. For given $i$ and $b$, $\lambda_{i,b}^{\overline{\mathrm{MAP}}}$, is obtained by traversing only those parts of the search tree that have leaves in $\mathcal{X}_{i,b}^{\left(x_{i,b}^{\overline{\mathrm{MAP}}}\right)}$. The quantities $\lambda^{\mathrm{MAP}}$ and $\lambda_{i,b}^{\overline{\mathrm{MAP}}}$ can, in principle, be computed using the sphere decoder based on the repeated tree-search (RTS) approach described in [19]. The RTS strategy results, however, in redundant computations as (often significant) parts of the search tree are revisited during the RTS steps required to determine $\lambda_{i,b}^{\overline{\mathrm{MAP}}}$ for all $i, b$. Following the STS paradigm described for soft-output SD in [7], we note that *efficient* computation of $L_{i,b}^{\mathrm{D}}$ ($\forall i, b$) requires that every node in the tree be visited *at most* once. This can be achieved by searching for the MAP solution and computing the metrics $\lambda_{i,b}^{\overline{\mathrm{MAP}}}$ ($\forall i, b$) concurrently while ensuring that the subtree originating from a given node in the tree is pruned if searching that subtree can not lead to an update of either $\lambda^{\mathrm{MAP}}$ or at least one of the $\lambda_{i,b}^{\overline{\mathrm{MAP}}}$. Besides extending the ideas in [7] to take into account a priori information, the main idea underlying SISO STS-SD presented in this paper is to *directly* compute the extrinsic LLRs $L_{i,b}^{\mathrm{E}}$ through a tree search, rather than computing $L_{i,b}^{\mathrm{D}}$ first and then evaluating $L_{i,b}^{\mathrm{E}} = L_{i,b}^{\mathrm{D}} - L_{i,b}^{\mathrm{A}}$ ($\forall i, b$).

Due to the large dynamic range of LLRs, fixed-point detector implementations need to constrain the magnitude of the LLR values. Evidently, clipping of the LLR magnitude leads to a performance degradation in terms of error rate. It was noted in [7], [26] that incorporating LLR clipping into the tree search is very effective in terms of reducing the complexity of max-log soft-output SD. In addition, as demonstrated in [7], LLR clipping (when built into the tree search) also allows to tune the MIMO detection algorithm in terms of complexity versus performance by adjusting the clipping parameter. In the SISO case, we are ultimately interested in the *extrinsic* LLRs $L_{i,b}^{\mathrm{E}}$ and clipping should therefore ensure that $\left| L_{i,b}^{\mathrm{E}} \right| \leq L_{\max}$ ($\forall i, b$), where $L_{\max}$ is the LLR clipping parameter. It is therefore sensible to ask whether clipping of the *extrinsic* LLRs can be built directly into the tree search. The answer is in the affirmative and the corresponding solution is described below. We start by writing the extrinsic LLRs as

$$L_{i,b}^{\mathrm{E}} = \begin{cases} \Lambda_{i,b}^{\overline{\mathrm{MAP}}} - \lambda^{\mathrm{MAP}}, & x_{i,b}^{\mathrm{MAP}} = +1 \\ \lambda^{\mathrm{MAP}} - \Lambda_{i,b}^{\overline{\mathrm{MAP}}}, & x_{i,b}^{\mathrm{MAP}} = -1 \end{cases} \tag{25}$$

where the quantities

$$\Lambda_{i,b}^{\overline{\mathrm{MAP}}} = \begin{cases} \lambda_{i,b}^{\overline{\mathrm{MAP}}} - L_{i,b}^{\mathrm{A}}, & x_{i,b}^{\mathrm{MAP}} = +1 \\ \lambda_{i,b}^{\overline{\mathrm{MAP}}} + L_{i,b}^{\mathrm{A}}, & x_{i,b}^{\mathrm{MAP}} = -1 \end{cases} \quad (26)$$

will be referred to as the *extrinsic metrics*. For the following developments it will be convenient to define the function $f(\cdot)$ that transforms an intrinsic metric $\lambda$ with associated a priori LLR $L^{\mathrm{A}}$ and binary label $x$ to an extrinsic metric $\Lambda$ according to

$$\Lambda = f\big(\lambda, L^{\mathrm{A}}, x\big) = \begin{cases} \lambda - L^{\mathrm{A}}, & x = +1 \\ \lambda + L^{\mathrm{A}}, & x = -1. \end{cases} \quad (27)$$

With this notation, we can rewrite (26) more compactly as $\Lambda_{i,b}^{\overline{\mathrm{MAP}}} = f\big(\lambda_{i,b}^{\overline{\mathrm{MAP}}}, L_{i,b}^{\mathrm{A}}, x_{i,b}^{\mathrm{MAP}}\big)$. The inverse function of (27) transforms an extrinsic metric $\Lambda$ to an intrinsic metric $\lambda$ and is given by

$$\lambda = f^{-1}\big(\Lambda, L^{\mathrm{A}}, x\big) = \begin{cases} \Lambda + L^{\mathrm{A}}, & x = +1 \\ \Lambda - L^{\mathrm{A}}, & x = -1. \end{cases} \quad (28)$$

We emphasize that the tree-search algorithm described in the following produces the extrinsic LLRs $L_{i,b}^{\mathrm{E}}$ ($\forall i, b$) in (25) rather than the intrinsic ones in (10). Since the soft-output STS-SD algorithm described in [7] delivers $L_{i,b}^{\mathrm{D}}$ and $L_{i,b}^{\mathrm{E}} = L_{i,b}^{\mathrm{D}}$ only in the soft-output case (i.e., if $L_{i,b}^{\mathrm{A}} = 0$, $\forall i, b$), careful modification of the list administration steps, the tree-pruning criterion, and the LLR clipping rules, of the soft-output STS-SD algorithm, is needed.

### A. List Administration

The main idea of the SISO STS paradigm is to search the subtree originating from a given node only if the result can lead to an update of either $\lambda^{\mathrm{MAP}}$ or of at least one of the $\Lambda_{i,b}^{\overline{\mathrm{MAP}}}$. To this end, the SD algorithm needs to maintain a list containing the current MAP hypothesis $\mathbf{x}^{\mathrm{MAP}}$, the corresponding metric $\lambda^{\mathrm{MAP}}$, and all $QM_{\mathrm{T}}$ extrinsic metrics $\Lambda_{i,b}^{\overline{\mathrm{MAP}}}$. The algorithm is initialized with $\lambda^{\mathrm{MAP}} = \Lambda_{i,b}^{\overline{\mathrm{MAP}}} = \infty$ and $x_{i,b}^{\mathrm{MAP}} = 1$ ($\forall i, b$). Whenever a leaf node with corresponding label $\mathbf{x}$ has been reached, the detector distinguishes between two cases:

*i) MAP hypothesis update:* If $d(\mathbf{x}) < \lambda^{\mathrm{MAP}}$, a new MAP hypothesis has been found. First, all extrinsic metrics $\Lambda_{i,b}^{\overline{\mathrm{MAP}}}$ for which $x_{i,b} = \overline{x_{i,b}^{\mathrm{MAP}}}$ are updated according to

$$\Lambda_{i,b}^{\overline{\mathrm{MAP}}} \leftarrow f\big(\lambda^{\mathrm{MAP}}, L_{i,b}^{\mathrm{A}}, x_{i,b}^{\overline{\mathrm{MAP}}}\big)$$

followed by the updates $\lambda^{\mathrm{MAP}} \leftarrow d(\mathbf{x})$ and $\mathbf{x}^{\mathrm{MAP}} \leftarrow \mathbf{x}$. In other words, for each bit in the MAP hypothesis that is changed in the update process, the metric associated with the *former* MAP hypothesis becomes the extrinsic metric of the *new* counter-hypothesis.

*ii) Extrinsic metric update:* In the case where $d(\mathbf{x}) > \lambda^{\mathrm{MAP}}$, only extrinsic metrics corresponding to counter-hypotheses might be updated. For each $i = 1, \ldots, M_{\mathrm{T}}$, $b = 1, \ldots, Q$ with $x_{i,b} = \overline{x_{i,b}^{\mathrm{MAP}}}$ and $f\big(d(\mathbf{x}), L_{i,b}^{\mathrm{A}}, x_{i,b}^{\mathrm{MAP}}\big) < \Lambda_{i,b}^{\overline{\mathrm{MAP}}}$, the SISO STS-SD algorithm performs the update

$$\Lambda_{i,b}^{\overline{\mathrm{MAP}}} \leftarrow f\big(d(\mathbf{x}), L_{i,b}^{\mathrm{A}}, x_{i,b}^{\mathrm{MAP}}\big) . \tag{29}$$

## B. Extrinsic LLR Clipping

In order to ensure that the extrinsic LLRs delivered by the algorithm indeed satisfy $\big| L_{i,b}^{E} \big| \leq L_{\max}$ $(\forall i, b)$, the following update rule

$$\Lambda_{i,b}^{\overline{\mathrm{MAP}}} \leftarrow \min\left\{ \Lambda_{i,b}^{\overline{\mathrm{MAP}}}, \lambda^{\mathrm{MAP}} + L_{\max} \right\}, \quad \forall i, b \tag{30}$$

has to be applied after carrying out the steps in Case i) of the list administration procedure described in Section III-A. Fig. 2 illustrates the principle of extrinsic LLR clipping. The search for counter-hypotheses associated with extrinsic metrics is constrained to a hypersphere of radius $r = \sqrt{\lambda^{\mathrm{MAP}} + L_{\max}}$ around the (transformed) received signal vector $\tilde{\mathbf{y}}$. In Section VI-B1, it will be demonstrated numerically that incorporating the constraint $|L_{i,b}^{\mathrm{E}}| \leq L_{\max}$ directly into the tree search significantly reduces complexity. We emphasize that for $L_{\max} = \infty$ the detector attains max-log optimal SISO performance, whereas for $L_{\max} = 0$, the LLRs satisfy $L_{i,b}^{\mathrm{E}} = 0$ and the hard-output MAP solution (8) is obtained.

## C. The Tree-Pruning Criterion

Consider the node $\mathbf{s}^{(i)}$ on level $i$ corresponding to the label bits $x_{j,b}$ $(j = i, \ldots, M_{\mathrm{T}}, b = 1, \ldots, Q)$. Assume that the subtree originating from this node and corresponding to the label bits $x_{j,b}$ $(j = 1, \ldots, i-1, b = 1, \ldots, Q)$ has not been expanded yet. The tree-pruning criterion for the node $\mathbf{s}^{(i)}$ along with its subtree is compiled from two sets, defined as follows:

1) The bits in the partial label $\mathbf{x}^{(i)}$ corresponding to the node $\mathbf{s}^{(i)}$ are compared with the corresponding bits in the label of the current MAP hypothesis. All extrinsic metrics $\Lambda_{i,b}^{\overline{\mathrm{MAP}}}$

with $x_{i,b} = \overline{x_{i,b}^{\mathrm{MAP}}}$ found in this comparison, may be affected when searching the subtree originating from $\mathbf{s}^{(i)}$. As $d(\mathbf{x}^{(i)})$ is an intrinsic metric, the extrinsic metrics $\Lambda_{i,b}^{\overline{\mathrm{MAP}}}$ need to be mapped to intrinsic metrics according to (28). The resulting set of *intrinsic* metrics, which may be affected by an update, is given by

$$\mathcal{A}_1\Big(\mathbf{x}^{(i)}\Big) = \Big\{ f^{-1}\Big(\Lambda_{j,b}^{\overline{\mathrm{MAP}}}, L_{j,b}^{\mathrm{A}}, x_{j,b}^{\mathrm{MAP}}\Big) \,\Big|\, \big(j \geq i, \forall b\big)$$
$$\wedge \Big(x_{j,b} = \overline{x_{j,b}^{\mathrm{MAP}}}\Big) \Big\}.$$

2) The extrinsic metrics $\Lambda_{j,b}^{\overline{\mathrm{MAP}}}$ for $j = 1, \ldots, i-1$, $b = 1, \ldots, Q$ corresponding to the counter-hypotheses in the subtree of $\mathbf{s}^{(i)}$ may be affected as well. Correspondingly, we define

$$\mathcal{A}_2\Big(\mathbf{x}^{(i)}\Big) = \Big\{ f^{-1}\Big(\Lambda_{j,b}^{\overline{\mathrm{MAP}}}, L_{j,b}^{\mathrm{A}}, x_{j,b}^{\mathrm{MAP}}\Big) \,\Big|\, j < i, \forall b \Big\}.$$

The intrinsic metrics which may be affected during the search in the subtree originating from node $\mathbf{s}^{(i)}$ are given by $\mathcal{A}(\mathbf{x}^{(i)}) = \{a_l\} = \mathcal{A}_1(\mathbf{x}^{(i)}) \cup \mathcal{A}_2(\mathbf{x}^{(i)})$. The node $\mathbf{s}^{(i)}$ along with its subtree is pruned if the corresponding PD $d(\mathbf{x}^{(i)})$ satisfies the tree-pruning criterion

$$d\big(\mathbf{x}^{(i)}\big) > \max_{a_l \in \mathcal{A}(\mathbf{x}^{(i)})} a_l.$$

This tree-pruning criterion ensures that a given node and the entire subtree originating from that node are explored only if this could lead to an update of either $\lambda^{\mathrm{MAP}}$ or of at least one of the extrinsic metrics $\Lambda_{i,b}^{\overline{\mathrm{MAP}}}$. Note that $\lambda^{\mathrm{MAP}}$ does not appear in the set $\mathcal{A}(\mathbf{x}^{(i)})$, as the update criteria given in Section III-A ensure that $\lambda^{\mathrm{MAP}}$ is always smaller than or equal to all intrinsic metrics associated with the counter-hypotheses.

## IV. CHANNEL-MATRIX PREPROCESSING

In this section, we describe how performing the QR-decomposition (QRD) on a column-sorted and regularized version of the channel matrix $\mathbf{H}$ in combination with compensation of self-interference in the LLRs —caused by channel-matrix regularization— carried out directly in the tree search can result in a significant complexity reduction at negligible performance loss. The use of column-sorting and regularization for soft-output SD was discussed in detail in [7]. We shall therefore keep the discussion of the general aspects short and emphasize the aspects corresponding to self-interference compensation.

### A. Column-Sorting and Regularization of the Channel Matrix

Methods for column-sorting and regularization of the channel matrix $\mathbf{H}$ performed on the basis of the received symbol vector $\mathbf{y}$ have been discussed, e.g., in [27], [28]. Unfortunately, such techniques require QRD on symbol-vector rate, which leads to a significant computational burden. In contrast, column-sorting and regularization based solely on the channel matrix $\mathbf{H}$ (and possibly on the noise variance) require QRDs only when the channel state changes, which entails a significantly smaller computational burden.

*1) Column-sorting:* The complexity of SD can be reduced (often significantly) by performing the QRD on a column-sorted version of $\mathbf{H}$ rather than on $\mathbf{H}$ directly, i.e., by computing $\mathbf{HP} = \mathbf{QR}$, where $\mathbf{P}$ is an $M_\mathrm{T} \times M_\mathrm{T}$ permutation matrix. Reduction in terms of complexity is obtained if levels closer to the root correspond to main-diagonal entries of $\mathbf{R}$ with larger magnitude, or equivalently, to spatial streams with higher effective SNR. A corresponding computationally efficient heuristic was proposed in [29] and is referred to as sorted QRD (SQRD) in the following.

*2) Regularization:* A further reduction in terms of complexity—at the cost of slightly reduced performance—can be obtained by performing the tree search on a Tikhonov-regularized (and column-sorted) version of $\mathbf{H}$ according to [30]

$$\underbrace{\begin{bmatrix} \mathbf{H} \\ \alpha \mathbf{I}_{M_\mathrm{T}} \end{bmatrix}}_{\overline{\mathbf{H}}} \mathbf{P} = \underbrace{\begin{bmatrix} \mathbf{Q}_a & \mathbf{Q}_c \\ \mathbf{Q}_b & \mathbf{Q}_d \end{bmatrix}}_{\overline{\mathbf{Q}}} \underbrace{\begin{bmatrix} \tilde{\mathbf{R}} \\ \mathbf{0}_{M_\mathrm{R} \times M_\mathrm{T}} \end{bmatrix}}_{\overline{\mathbf{R}}} \tag{31}$$

where $\alpha \in \mathbb{R}$ is a suitably chosen regularization parameter. Here, $\overline{\mathbf{R}}$ and $\overline{\mathbf{Q}}$ are partitioned such that $\tilde{\mathbf{R}}$, $\mathbf{Q}_a$, $\mathbf{Q}_b$, $\mathbf{Q}_c$, and $\mathbf{Q}_d$ are of dimension $M_\mathrm{T} \times M_\mathrm{T}$, $M_\mathrm{R} \times M_\mathrm{T}$, $M_\mathrm{T} \times M_\mathrm{T}$, $M_\mathrm{R} \times M_\mathrm{R}$, and $M_\mathrm{T} \times M_\mathrm{R}$, respectively. The computational complexity for regularized SQRD as compared to non-regularized SQRD is approximately 50% higher [31]. However, the QRD needs to be performed only if the channel matrix $\mathbf{H}$ changes, as opposed to the tree-search itself, which needs to be carried out at symbol-vector rate.

LLR computation (and MAP detection) based on regularized SQRD corresponds to replacing the modified input-output relation in (4) by

$$\hat{\mathbf{y}} = \tilde{\mathbf{R}}\tilde{\mathbf{s}} + \tilde{\mathbf{n}} \tag{32}$$

where $\hat{\mathbf{y}} = \mathbf{Q}_a^H \mathbf{y}$, $\tilde{\mathbf{s}} = \mathbf{P}^T \mathbf{s}$, and $\tilde{\mathbf{n}} = -\alpha \mathbf{Q}_b^H \mathbf{s} + \mathbf{Q}_a^H \mathbf{n}$. The corresponding intrinsic (max-log) LLRs in (5) are obtained by pretending that the resulting noise $\tilde{\mathbf{n}}$ has the same statistics as $\mathbf{n}$, which leads to

$$\tilde{L}_{i,b}^{\mathrm{D}} \triangleq \min_{\tilde{\mathbf{s}} \in \mathcal{X}_{i,b}^{(-1)}} \left\{ \frac{1}{N_{\mathrm{o}}} \left\| \hat{\mathbf{y}} - \tilde{\mathbf{R}} \tilde{\mathbf{s}} \right\|^2 - \log \mathrm{P}[\tilde{\mathbf{s}}] \right\}$$
$$- \min_{\tilde{\mathbf{s}} \in \mathcal{X}_{i,b}^{(+1)}} \left\{ \frac{1}{N_{\mathrm{o}}} \| \hat{\mathbf{y}} - \tilde{\mathbf{R}} \tilde{\mathbf{s}} \|^2 - \log \mathrm{P}[\tilde{\mathbf{s}}] \right\} \tag{33}$$

where $\mathrm{P}[\tilde{\mathbf{s}}] = \mathrm{P}[\mathbf{s}]$. The intrinsic LLRs $\tilde{L}_{i,b}^{\mathrm{D}}$ in (33) will, in general, only be approximations to the true intrinsic LLRs $L_{i,b}^{\mathrm{D}}$ in (5). This is a consequence of $\tilde{\mathbf{n}}$ no longer being i.i.d. circularly symmetric complex Gaussian distributed with variance $N_{\mathrm{o}}$ per complex entry, as it contains self-interference (i.e., it depends on $\mathbf{s}$) and $\mathbf{Q}_a$ is, in general, not unitary. Setting $\alpha = \sqrt{N_{\mathrm{o}}/\mathbb{E}[|s|^2]}$, where we note that $\mathbb{E}[|s_i|^2] = \mathbb{E}[|s|^2], \forall i$, leads to the so-called minimum mean-square error (MMSE) SQRD [32] and has been shown in [7] to result in a good performance/complexity tradeoff for soft-output STS-SD. In the remainder of this paper, regularization will always refer to using MMSE-SQRD. Finally, we note that the LLRs in (33) need to be reordered after the detection stage to account for the permutation induced by $\mathbf{P}$.

## B. Compensation of Self-Interference

Using the approximate (max-log) LLRs in (33) with $\alpha \neq 0$ instead of the exact max-log LLRs in (5) results in a performance loss. In order to recover part of this performance loss, a method for the compensation of self-interference was developed in [33] for list-based MIMO detectors. The approach described in [33] can not be applied directly to SISO STS-SD. It turns out, however, that compensation of self-interference can be incorporated directly into the tree-search procedure. This leads to a noticeable performance improvement compared to using (33), while the corresponding increase in complexity is negligible (corresponding simulation results are shown in Section VI-B3).

*1) Compensation of self-interference:* As shown in [33], the squared Euclidean distance $\left\| \underline{\mathbf{y}} - \underline{\mathbf{H}}\mathbf{s} \right\|^2$ with $\underline{\mathbf{y}} = \begin{bmatrix} \mathbf{y}^T & \mathbf{0}_{1 \times M_T} \end{bmatrix}^T$ can be expanded in two different ways according to

$$\left\| \underline{\mathbf{y}} - \underline{\mathbf{H}}\mathbf{s} \right\|^2 = \| \mathbf{y} - \mathbf{H}\mathbf{s} \|^2 + \alpha^2 \| \mathbf{s} \|^2 \tag{34}$$

$$\left\| \underline{\mathbf{y}} - \underline{\mathbf{H}}\mathbf{s} \right\|^2 = \left\| \underline{\mathbf{Q}}^H \underline{\mathbf{y}} - \underline{\mathbf{R}}\mathbf{P}^T\mathbf{s} \right\|^2 = \left\| \hat{\mathbf{y}} - \tilde{\mathbf{R}}\tilde{\mathbf{s}} \right\|^2 + \left\| \mathbf{Q}_c^H \mathbf{y} \right\|^2 \tag{35}$$

where (35) is obtained by using (31). Equating the RHS terms of (34) and (35) and using $\|\mathbf{s}\|^2 = \|\tilde{\mathbf{s}}\|^2$ yields

$$\|\mathbf{y} - \mathbf{H}\mathbf{s}\|^2 = \left\|\hat{\mathbf{y}} - \tilde{\mathbf{R}}\tilde{\mathbf{s}}\right\|^2 + \left\|\mathbf{Q}_c^H \mathbf{y}\right\|^2 - \alpha^2 \|\tilde{\mathbf{s}}\|^2 \qquad (36)$$

which allows us to conclude that the metric $\|\mathbf{y} - \mathbf{H}\mathbf{s}\|^2$ contains a contribution that is independent of the symbol vectors, namely $\|\mathbf{Q}_c^H \mathbf{y}\|^2$, and a term caused by self-interference given by $-\alpha^2 \|\tilde{\mathbf{s}}\|^2$. Since we use $\left\|\hat{\mathbf{y}} - \tilde{\mathbf{R}}\tilde{\mathbf{s}}\right\|^2$ (instead of the left-hand side of (36)) in the LLR computation (33), the two remaining RHS-terms in (36) must be compensated. As already observed in Section II-C, constant terms (i.e., terms that are independent of $\mathbf{s}$) cancel out in the LLR computation (10) and can therefore be neglected without affecting the resulting LLRs, whereas the term $-\alpha^2 \|\tilde{\mathbf{s}}\|^2$ does depend on $\mathbf{s}$ and therefore needs to be compensated. This is accomplished by computing the *self-interference free* (SIF) intrinsic max-log LLRs according to [33]

$$\bar{L}_{i,b}^{\mathrm{D}} \triangleq \min_{\tilde{\mathbf{s}} \in \mathcal{X}_{i,b}^{(-1)}} \left\{ \frac{1}{N_{\mathrm{o}}} \left\|\hat{\mathbf{y}} - \tilde{\mathbf{R}}\tilde{\mathbf{s}}\right\|^2 - \frac{\alpha^2}{N_{\mathrm{o}}} \|\tilde{\mathbf{s}}\|^2 - \log \mathrm{P}[\tilde{\mathbf{s}}] \right\}$$

$$- \min_{\tilde{\mathbf{s}} \in \mathcal{X}_{i,b}^{(+1)}} \left\{ \frac{1}{N_{\mathrm{o}}} \|\hat{\mathbf{y}} - \tilde{\mathbf{R}}\tilde{\mathbf{s}}\|^2 - \frac{\alpha^2}{N_{\mathrm{o}}} \|\tilde{\mathbf{s}}\|^2 - \log \mathrm{P}[\tilde{\mathbf{s}}] \right\}. \qquad (37)$$

We emphasize, however, that (37) remains an approximation to (5) as the noise term $\tilde{\mathbf{n}}$ resulting from (32) is *not* i.i.d. circularly symmetric Gaussian distributed with variance $N_{\mathrm{o}}$ per complex entry.

*2) Compensation in the SISO STS-SD algorithm:* In [33] it was suggested to compensate self-interference *after* the tree-search. For the SISO STS-SD algorithm, extrinsic LLRs are computed *only* on the basis of the MAP hypothesis $\mathbf{x}^{\mathrm{MAP}}$, its metric $\lambda^{\mathrm{MAP}}$, and extrinsic metrics $\Lambda_{i,b}^{\overline{\mathrm{MAP}}}$ (see (25)). Compensation of self-interference according to (37) *after* carrying out the SISO STS-SD algorithm, would additionally require explicit knowledge of the symbol vectors $\mathbf{s} \in \mathcal{X}_{i,b}^{\left(\overline{x_{i,b}^{\mathrm{MAP}}}\right)}$, which is, in general, not available. Inspection of (37) suggests, however, that self-interference compensation may be incorporated into the tree-search procedure. Straightforward modification of the DIs in (23) to accomplish this would lead to the modified DIs

$$|\tilde{e}_i| = |e_i| - \frac{\alpha^2}{N_{\mathrm{o}}} |\tilde{s}_i|^2$$

which are, however, no longer guaranteed to be strictly non-negative. As in the tightening of the tree-pruning criterion described in Section II-B, we recognize that symbol-vector-independent terms can be added to the DIs without loss of (max-log) optimality. Therefore, setting the DIs to

$$|\bar{e}_i| \triangleq |e_i| + m(\tilde{s}_i) \tag{38}$$

with the non-negative term

$$m(\tilde{s}_i) = \frac{\alpha^2}{N_o} \left( \max_{s \in \mathcal{O}} |s|^2 - |\tilde{s}_i|^2 \right) \tag{39}$$

leads to the smallest possible non-negative DIs that compensate self-interference directly in the tree search. Note that adding non-negative terms to the DIs as done in (38), in general, increases the (tree-search) complexity. Recall, however, that channel-matrix regularization itself almost always significantly reduces complexity [7], so that this increase, which is shown numerically in Section VI-B3 to be marginal, is tolerable. In addition, it turns out that self-interference compensation recovers the performance loss due to channel-matrix regularization to a point where near-max-log optimal performance is achieved (see Section VI-B3). In the case of constant-modulus symbol alphabets (e.g., BPSK or 4-QAM) we have $m(\tilde{s}_i) = 0$ ($i = 1, \ldots, M_T$) and compensation of self-interference in the tree-search does not even increase complexity. We conclude by noting that the quantities $\max_{s_i \in \mathcal{O}} |s_i|^2$ can be pre-computed and hence, the additional computational complexity required to incorporate compensation of self-interference into the tree-search procedure is small.

## V. LLR CORRECTION

The max-log approximation, channel-matrix regularization, and other complexity-reducing mechanisms, such as early termination of the tree-search [7], lead to LLRs that are approximations to the true LLRs in (2). However, channel decoders (see Fig. 1) rely on exact LLRs to achieve optimum performance. In the following, we present a post-processing method for correcting approximate LLRs resulting from sub-optimum detectors. This method is based on ideas developed in [34] and [35] and is able to (often significantly) improve the performance in (iterative) MIMO decoders while requiring low additional computational complexity.

### A. The Basic Idea

We start by defining (or recalling the definitions of) the following objects (see Fig. 3):

- the *effective channel* with the binary-valued inputs $x_{i,b}$ and the associated a priori LLRs $L_{i,b}^{\mathrm{A}}$ and outputs given by the (possibly approximated) extrinsic LLRs $L_{i,b}^{\mathrm{E}}$.

- the *physical MIMO channel* with input $\mathbf{s}$ and output $\mathbf{y}$.

- the *soft-input soft-output MIMO detector* with inputs $\mathbf{y}$ and $L_{i,b}^{\mathrm{A}}$ and outputs $L_{i,b}^{\mathrm{E}}$.

- the *LLR correction unit* (see Fig. 3) computes corrected extrinsic LLRs $L_{i,b}^{\mathrm{C}}$ based on (approximated) extrinsic LLRs $L_{i,b}^{\mathrm{E}}$ and on side information $\mathcal{Z}$, by applying an LLR correction function

$$L_{i,b}^{\mathrm{C}} = g\left(L_{i,b}^{\mathrm{E}}, \mathcal{Z}\right). \tag{40}$$

- the *side information* $\mathcal{Z}$ is, for example, obtained from the (instantaneous) receive SNR, the singular values of the channel matrix $\mathbf{H}$, and from knowledge of whether the soft-input soft-output MIMO detector was terminated prematurely [7].

For the LLR correction function to yield valid LLRs, we define

$$g\left(L_{i,b}^{\mathrm{E}}, \mathcal{Z}\right) = \log\left(\frac{\mathrm{P}\left[x_{i,b} = +1 \mid L_{i,b}^{\mathrm{E}}, \mathcal{Z}\right]}{\mathrm{P}\left[x_{i,b} = -1 \mid L_{i,b}^{\mathrm{E}}, \mathcal{Z}\right]}\right). \tag{41}$$

Just like the LLRs in (2) are computed based on the received vector $\mathbf{y}$ and the channel state $\mathbf{H}$, the corrected LLRs are computed based on the (approximated) extrinsic LLRs $L_{i,b}^{\mathrm{E}}$ and the side information $\mathcal{Z}$. The formulation (40) and (41) entails that $L_{i,b}^{\mathrm{C}}$ depends only on $L_{i,b}^{\mathrm{E}}$ (and $\mathcal{Z}$) rather than on all extrinsic (approximated) LLR values $L_{i,b}^{\mathrm{E}}$ (for all $i, b$). Making the correction function depend on other LLR values, besides the one to be corrected, would certainly improve the correction performance, but at the same time also dramatically increase the computational effort for LLR correction, as will become clear in the discussion of the numerical procedure for LLR correction proposed below.

The main idea is now, depending on the mechanisms used to approximate the extrinsic LLRs (e.g., the max-log approximation, channel-matrix regularization, early termination of the tree-search), to extract suitable side information $\mathcal{Z}$. To see that this is non-trivial and the problem is multi-faceted, simply note that the set of all possible channel matrices $\mathbf{H}$ is a continuum of $M_{\mathrm{R}} \times M_{\mathrm{T}}$ complex-valued matrices. This continuum will be absorbed in $\mathcal{Z}$ through, e.g., the singular values or the rank of $\mathbf{H}$. We emphasize that in practice, LLR correction requires that the set $\mathcal{Z}$ be finite. In addition, the individual entries of $\mathcal{Z}$ must have finite cardinality as well. Hence, continuous-valued quantities, such as, e.g., the SNR or singular values, must be suitably quantized. The total number of different instances of the side information $\mathcal{Z}$ is denoted by $Z$ in the following.

### B. Computation of the LLR Correction Function

Once we have chosen $\mathcal{Z}$, the LLR correction function (41) is —in principle— obtained from the conditional probabilities $\mathrm{P}\big[x_{i,b} = \pm 1 \,|\, L_{i,b}^{\mathrm{E}}, \mathcal{Z}\big]$. Analytical expressions for correction functions seem very hard to obtain (even for simple examples such as for Hagenauer's approximation to the box function [34]). We next propose an approach for numerically computing (approximations to) the LLR correction function in (41).

First, the range of the LLRs to be corrected needs to be constrained (motivated, e.g., by the use of LLR clipping) to $L_{i,b}^{\mathrm{E}} \in [-L_{\max}, +L_{\max}]$. This interval is then divided into $K$ equally-sized bins such that the $k$th bin corresponds to

$$\mathcal{B}_k = \left[ -L_{\max} + k\frac{2L_{\max}}{K}, -L_{\max} + (k+1)\frac{2L_{\max}}{K} \right), \quad k = 0, \dots, K-1.$$

Then, the histogram

$$p_k(\mathcal{Z}) = \mathrm{P}\big[x_{i,b} = +1 \,|\, L_{i,b}^{\mathrm{E}} \in \mathcal{B}_k, \mathcal{Z}\big], \quad k = 0, \dots, K-1 \tag{42}$$

can be computed by performing Monte-Carlo simulations (averaged over noise and channel realizations) with randomly generated bits $x_{i,b}$. For each $L_{i,b}^{\mathrm{E}}$ and a given instance of $\mathcal{Z}$, the (approximated) LLR correction function is obtained by linear interpolation between the base points

$$\left( -L_{\max} + \left(k + \frac{1}{2}\right)\frac{2L_{\max}}{K} \,,\, \log\left(\frac{p_k(\mathcal{Z})}{1 - p_k(\mathcal{Z})}\right) \right). \tag{43}$$

We emphasize that for each instance of $\mathcal{Z}$, in general, a different LLR correction function is obtained. Note that the LLRs resulting from (43) can have a magnitude that is larger than $L_{\max}$ (see Section VI-D). The corrected LLRs can be clipped again to satisfy $\big|L_{i,b}^{\mathrm{C}}\big| \leq L_{\max,\mathrm{c}}$, where $L_{\max,\mathrm{c}} \geq L_{\max}$, thereby limiting the dynamic range of LLRs (rather than performing LLR clipping for complexity reduction and tuning of the detector as done so far).

The computational complexity needed to compute the histogram (42) and the corresponding storage requirements depend critically on the number of bins $K$ and on the total number of different instances of the side information $\mathcal{Z}$ given by $Z$. In particular, $ZK$ histogram values need to be stored and hence, it is important to keep both $Z$ and $K$ small. Application of the LLR correction function itself amounts to simple table look-up operations followed by linear interpolation, which can be performed at very low computational complexity.

## C. An Example

We next discuss an example that illustrates the impact (and importance) of LLR correction. The complexity of the SISO STS-SD algorithm depends critically on the noise realization $\mathbf{n}$, the channel-matrix realization $\mathbf{H}$, the transmit-vector $\mathbf{s}$, and the a priori LLRs $L_{i,b}^{\mathrm{A}}$. The often prohibitively high worst-case complexity of SD [36] constitutes a problem in many practical application scenarios, as it inhibits realizing the throughput requirements of many of the available communication standards. A promising approach to limiting the worst-case complexity of SD, while keeping the resulting performance degradation small, was proposed in [37], [7]. The basic idea is to impose an aggregate complexity constraint of $ND_{\mathrm{avg}}$ visited nodes for a block of $N$ symbol vectors by using maximum-first (MF) scheduling. This scheduling strategy allocates the overall complexity budget according to

$$D_{\max}[j] = ND_{\mathrm{avg}} - \sum_{\ell=1}^{j-1} D[\ell] - (N-j)M, \quad j = 1, \ldots, N \qquad (44)$$

where $M$ is a parameter to be specified below and $D[\ell]$ is the actual number of nodes visited in the detection of the $\ell$th symbol vector with a corresponding maximum complexity constraint of $D_{\max}[j]$, i.e., the detector is terminated if $D_{\max}[j]$ nodes have been visited and the LLRs are obtained from the current MAP hypothesis, the associated metric $\lambda^{\mathrm{MAP}}$, and the current extrinsic metrics $\Lambda_{i,b}^{\overline{\mathrm{MAP}}}$. The main idea realized by the policy (44) is that detection of the $j$th symbol vector is allowed to use up all of the remaining complexity budget within the block of $N$ symbol vectors up to $(N-j)M$ nodes, i.e., the parameter $M$ determines that in decoding the remaining $N-j$ symbol vectors, we can afford a budget of at least $M$ nodes per symbol vector. Setting $M = M_{\mathrm{T}}$ and choosing $D_{\mathrm{avg}} \geq M_{\mathrm{T}}$ (what is used in the remainder of the paper), ensures that for each of the remaining $N-j$ symbol vectors at least the hard-output successive interference cancellation (SIC) solution is found [7]. For details on early termination and scheduling, we refer to [37], [7].

Now, if early termination happens before the extrinsic metric $\Lambda_{i,b}^{\overline{\mathrm{MAP}}}$ was updated from its initial value $\infty$, the corresponding LLR satisfies $\left| L_{i,b}^{\mathrm{E}} \right| = L_{\max}$ as only LLR clipping according to (30) was performed. Hence, early termination may result in LLRs with a higher reliability than they would actually have if no complexity constraints were imposed. This calls for LLR correction with the goal of reducing the magnitude of such LLRs. Consequently, the side information set $\mathcal{Z}$ should contain a binary-valued state variable, which indicates whether early termination occurred or not. Corresponding numerical results are provided in Section VI-D1.

# VI. SIMULATION RESULTS

Unless explicitly stated otherwise, all simulation results are for a convolutionally encoded (rate $R = 1/2$, generator polynomials [$133_o$ $171_o$], and constraint length 7) iterative MIMO-OFDM system with $M_T = M_R = 4$, 16-QAM constellation $\mathcal{O}$ with Gray labeling, 64 OFDM tones, TGn type C channel model [38], and are based on a max-log BCJR channel decoder [39]. One frame consists of $1024$ randomly interleaved (across space and frequency) bits corresponding to one (spatial) OFDM symbol and we assume that the bits $x_{i,b}$ ($\forall i, b$) are statistically independent. The SNR is per receive antenna and the SNR values specified in the figures are in decibels (dBs). The number of iterations $I$ is the number of times the soft-input soft-output MIMO detector (and the SISO channel decoder) are used, i.e., $I = 1$ corresponds to soft-output SD in [7]. The LLR clipping parameters shown in the simulation results correspond to *normalized* LLR clipping parameters according to $L_{\max}/N_o$.

## A. Tightening of the Tree-Pruning Criterion

*1) Impact of the Euclidean-distance term:* The goal of the simulation results shown in Table I is to quantify the impact of the Euclidean distance term $\frac{1}{N_o} \left| \tilde{y}_i - \sum_{j=i}^{M_T} R_{i,j} s_j \right|^2$ in the bias (15) on the complexity reduction obtained by tightening the tree-pruning criterion according to (17). To this end, we set the prior term to zero, i.e., $\log P[\mathbf{s}] = 0$, and compare the complexity resulting from the tightened tree-pruning criterion to that of the standard tree-pruning criterion (denoted by "std." in Table I) given in (14). We observe that the complexity reduction obtained by tightening of the tree-pruning criterion based on the Euclidean distance-term only, is marginal, in particular in the light of the prohibitive effort required to compute (15).

*2) Impact of the prior term:* Next, we start with uniform priors, i.e., $L_{i,b}^A = 0$ ($\forall i, b$) for the first iteration, and perform tightening of the tree-pruning criterion according to (19). Table II shows that removing the bias $|p_i|$ in (18) leads to a dramatic reduction in terms of complexity, ranging from 65.9% to 99.5%. Furthermore, we can see that the impact on complexity reduction in the second iteration ($I = 2$) is less pronounced (but still significant) than in the first iteration. This behavior can be explained by noting that for $I = 1$ the priors satisfy $L_{i,b}^A = 0$, which leads to the largest possible values for $|p_i|$, $i = 1, \ldots, M_T$. We note that, in general, the impact on complexity reduction is further reduced with increasing $I$.

We can now conclude that removing the Euclidean-distance component of the bias term (15) is not worth the effort. In contrast, tightening of the tree-pruning criterion based on the prior only (19) leads to a significant complexity reduction and requires no additional computational complexity if the individual bits $x_{i,b}$ ($\forall i, b$) are statistically independent (see (24) in Section II-C). In the remainder of this paper, we always employ tightening of the tree-pruning criterion according to (19).

## B. Performance/Complexity Tradeoffs

The performance/complexity tradeoffs discussed next and quantified in Figs. $4-6$, 8, and 9 refer to the *cumulative* (tree-search) complexity in terms of the total number of nodes visited (averaged over independent channel, noise, and data realizations) for SISO detection over $I$ iterations, designated as "average complexity" from now on. The computational complexity incurred by channel decoding is ignored in the following. The minimum SNR required to achieve a given frame error rate (FER) is referred to as the "SNR operating point" for that FER.

*1) Impact of LLR clipping:* From Fig. 4, we can conclude that LLR clipping allows for a smooth performance/complexity tradeoff, adjustable through a single parameter, namely the LLR clipping parameter $L_{\max}$. Note that for a fixed SNR operating point, the minimum complexity is not necessarily achieved by maximizing the number of iterations. The performance corresponding to the case where clipping of the extrinsic LLRs is performed *after* the tree search, i.e., LLR clipping is not incorporated into the tree search, is that obtained for $L_{\max} = \infty$. We can therefore conclude that incorporating LLR clipping into the tree search is of paramount importance as it reduces the complexity substantially and renders the detector easily adjustable in terms of performance versus complexity.

*2) Column-sorting and regularization:* We next examine the impact of column-sorting and regularization of the channel matrix on the performance/complexity tradeoff. It can be seen in Fig. 5 that in the low-complexity regime, the Pareto-optimal tradeoff curve is achieved by MMSE-SQRD. In the high-complexity regime, the performance loss incurred by regularization renders MMSE-SQRD inferior to un-regularized SQRD. This observation has already been made for the soft-output-only case in [7], but is also valid for $I > 1$ using SISO STS-SD.

*3) Self-interference free LLRs:* Fig. 5 also quantifies the impact of compensating self-interference — according to Section IV-B2 — on the performance/complexity tradeoff. We observe that compensation of self-interference results in a performance improvement in terms of SNR operating point of $0.3\,\mathrm{dB}$ to $0.5\,\mathrm{dB}$ in almost all regions. In the high-complexity regime un-regularized SQRD outperforms channel-matrix regularization and has an SNR operating point that is $0.15\,\mathrm{dB}$ below that obtained in the SIF case.

## C. Comparison with List Sphere Decoding

Fig. 6 compares the performance/complexity tradeoff achieved by list sphere decoding (LSD) as proposed in [1] to that obtained through SISO STS-SD. For the LSD algorithm, we take the complexity to equal the number of nodes visited when building the initial candidate list. The (often significant) computational burden incurred by list administration in LSD is neglected, leading to a complexity measure that favors the LSD algorithm. We can draw the following conclusions from Fig. 6:

   i) SISO STS-SD outperforms LSD for all SNR operating points.

  ii) LSD requires relatively large list sizes and hence a large amount of memory to approach (max-log) optimum SISO performance.[5] The underlying reason is that LSD obtains extrinsic LLRs from a candidate list that has been computed around the maximum-likelihood solution, i.e., in the absence of a priori information. In contrast, SISO STS-SD requires memory mainly for the extrinsic metrics, which are obtained through a search that is concentrated around the MAP solution. Consequently, SISO STS-SD tends to require (often significantly) less memory than LSD.

Besides LSD, various other SISO detection algorithms for MIMO systems have been developed, see e.g., [2]–[6], [41]. The algorithms described in [3] and [6] are related to LSD but require rebuilding the candidate list in each iteration; this can lead to a substantial complexity increase compared to LSD. For [2], [4] issues indicating potentially high computational complexity include the requirement for multiple matrix inversions for each symbol vector in each iteration.[6] In contrast, the QRD required for SD has to be computed only when the channel state changes. The computational complexity of the list-sequential (LISS) algorithm in [5], [41] seems difficult to relate to the complexity measure employed in this paper. However, due to the need for sorting of candidate vectors and the structural similarity of the LISS algorithm to LSD, we expect the performance/complexity tradeoff realized by the LISS algorithm to be comparable to that of the LSD algorithm.

---

[5]In addition to the memory requirements, the search-and-replace operations required in the LSD algorithm's list administration, quickly lead to prohibitively high VLSI implementation complexity when the list size grows [40].

### D. Impact of LLR Correction

Fig. 7 shows examples for LLR correction functions of SISO STS-SD obtained by linear interpolation using $K = 31$ bins and side information given by

$$\mathcal{Z} = \big\{ L_{\max}, D_{\mathrm{avg}}, \mathsf{SNR}, T \big\} \tag{45}$$

where $L_{\max} = 0.2$, $D_{\mathrm{avg}} \in \{16, \infty\}$, $\mathsf{SNR} = 16\,\mathrm{dB}$, and $T \in \{0, 1\}$ indicates whether early termination occurred ($T = 1$) or not ($T = 0$). Here, the number of instances of $\mathcal{Z}$ is given by $Z = 4$. Note that in practice, the parameters $L_{\max}$, $D_{\mathrm{avg}}$, and $\mathsf{SNR}$ in $\mathcal{Z}$ remain constant as long as the channel state remains constant, whereas $T$ may change at symbol-vector rate, i.e., depending on $T$, different LLR correction functions need to be applied to the extrinsic LLRs $L_{i,b}^{\mathrm{E}}$. We compare the LLR correction functions corresponding to SISO STS-SD using column-sorting (SQRD), regularization and column-sorting (MMSE-SQRD), and compensation of self-interference in combination with MMSE-SQRD, all having unconstrained maximum complexity (i.e., $D_{\mathrm{avg}} = \infty$ and, hence, $T = 0$). We furthermore show the correction function of SIF (MMSE-SQRD) LLRs in combination with MF scheduling for $D_{\mathrm{avg}} = 16$ (denoted by "MF16" in Fig. 7) and $T = 1$. The following observations can be made:

i) For unconstrained complexity, i.e., $D_{\mathrm{avg}} = \infty$, LLRs corresponding to $\pm L_{\max}$ are corrected to LLRs with larger magnitude; this is a result of clipping LLRs with magnitude larger than $L_{\max}$ to $\pm L_{\max}$. We note that since the LLR correction functions are obtained by binning and linear interpolation, LLR-values that have slightly smaller (mandated by the bin-width) magnitude than $L_{\max}$ are also corrected to values larger than $L_{\max}$.

---

[6]A detailed complexity analysis of the algorithm described in [2] based on VLSI implementation results can be found in [12].

ii) For early termination with MF-scheduling (i.e., $D_{\text{avg}} = 16$ and $T = 1$), LLRs with magnitude close to $L_{\max}$ are corrected to LLRs with smaller magnitude (i.e., their reliability is reduced). LLRs corresponding to $L_{i,b}^{\text{E}} = \pm L_{\max}$ are, as already mentioned in Section V-C, often caused by early termination and hence, are corrected to less reliable LLR-values.

iii) The LLR correction function associated with column-sorting (SQRD) only is almost a linear function with slope one, i.e., $L_{i,b}^{\text{C}} = L_{i,b}^{\text{E}}$, which indicates that little correction is performed. The reason for this behavior is that column-sorting maintains (max-log) optimality and the impact of the max-log approximation on performance is small, in general [12]. The correction functions associated with channel-matrix regularization show a stronger deviation from $L_{i,b}^{\text{C}} = L_{i,b}^{\text{E}}$ (cf. the zoom in Fig. 7), indicating that more correction is required, since regularization leads to an approximation of the max-log LLRs (see Section IV-A2).

*1) Performance/complexity tradeoff for SISO STS-SD with early termination:* Fig. 8 shows the performance/complexity tradeoff for early-termination based on MF scheduling with and without LLR correction. The side information was chosen according to (45) and the LLR correction function was computed based on $K = 31$ bins with linear interpolation. Depending on the average run-time constraint, LLR correction can reduce (i.e., improve) the SNR operating point by up to $3\,\text{dB}$. As expected, the performance gains resulting from LLR correction are more pronounced for larger clipping parameters as in these cases performance is dominated by the run-time constraint and early termination happens more often. Note that LLR correction also yields slight performance gains for small LLR clipping levels, where the run-time constraints do not affect performance. This indicates that LLR correction can also correct —at least partly— the errors induced by LLR clipping and by channel-matrix regularization.

*2) Performance/complexity tradeoff for parallel concatenated turbo codes:* The next simulation result is aimed at understanding which of the conclusions drawn so far change in the presence of more sophisticated channel codes. To this end, we evaluated the performance/complexity tradeoff for a parallel-concatenated turbo code (PCTC) of rate 1/2 (punctured, memory 2, and generator polynomial $[7_o\ 5_o]$, where $7_o$ pertains to the feedback path) with eight iterations in the turbo decoder. We use the interleaver specified in the 3GPP standard [42] with 508 information bits. One code-block corresponds to 1024 coded bits including two times four bits for termination of the trellises. For aggressive LLR clipping, simulation results have shown that using the sum-product algorithm within the turbo decoder requires precise (and hence, corrected) LLRs to yield satisfactory results, whereas max-log-based decoders seem to be more robust to effects incurred by LLR clipping [12]. Since we employ the sum-product BCJR algorithm [39] for decoding of the PCTCs, LLR correction is used.

The results in Fig. 9 indicate that the performance/complexity tradeoff achieved by the PCTC in the first iteration is significantly better than that obtained for the convolutional code (CC) used in the previous simulations. In the second iteration, the performance/complexity tradeoff is almost identical for both codes. For $I > 2$, the CC slightly outperforms the PCTC, which could be due to the fact that we use a turbo code with very short block length and a channel model that exhibits correlation across frequency and space (see, e.g., [43]).

### E. Information Transfer Characteristics

In order to characterize the performance of soft-input soft-output MIMO detectors *independently* of the channel code and channel decoder, we compute information transfer characteristics (ICTs) using an i.i.d. (across space and OFDM tones) Rayleigh multi-path fading channel model and assuming a Gaussian model for the a priori LLRs according to [44]

$$L_{i,b}^{\mathrm{A}} = \frac{2}{\sigma^2}\big(x_{i,b} + n\big)$$

where $n$ is a real-valued Gaussian RV with zero mean and variance $\sigma^2$. The a priori information content is determined by $\sigma^2$ and characterized by the mutual information between the transmitted bits $x_{i,b}$ and the a priori input of the SISO detector, i.e., $I_{\mathrm{A}} = I\big(x_{i,b}; L_{i,b}^{\mathrm{A}}\big)$ (in bits per binary symbol) where $0 \leq I_{\mathrm{A}} \leq 1$. Note that large and small values of $\sigma^2$, reduce and increase the mutual information $I_{\mathrm{A}}$, respectively. The extrinsic information at the output of the detector (averaged over all transmit antennas and bits) is defined as

$$I_{\mathrm{E}} = \frac{1}{M_{\mathrm{T}}Q} \sum_{i=1}^{M_{\mathrm{T}}} \sum_{b=1}^{Q} I\big(x_{i,b}; L_{i,b}^{\mathrm{E}}\big)$$

in bits per binary symbol where $0 \leq I_{\mathrm{E}} \leq 1$. Note that $L_{i,b}^{\mathrm{A}} = 0$ implies $I_{\mathrm{A}} = 0$ and corresponds to soft-output-only MIMO detection. The ITC corresponds to the function $I_{\mathrm{E}} = h(I_{\mathrm{A}})$, for a given SNR, and enables us to assess the performance of soft-input soft-output MIMO detectors in a fundamental way. Note that the application of ITCs as described here was originally proposed in [45, Chapter 16].

*1) Impact of LLR clipping:* Fig. 10 shows that a normalized LLR clipping parameter of $L_{max} = 0.4$ achieves almost the same ITC as max-log optimal SISO STS-SD with $L_{max} = \infty$. Hence, increasing the LLR clipping parameter to a value above 0.4 does not further improve performance of the detector and only leads to an increase in complexity. We note that the same observation was made in the performance/complexity tradeoff simulations in Fig. 4.

*2) Performance comparison with LSD:* Fig. 11 compares the ITC of SISO STS-SD to that of LSD [1]. For $I_A$ close to 1, LSD requires large list-sizes to yield a performance close to that of the max-log-optimal SISO STS-SD algorithm. Note that even hard-output MAP detection (which corresponds to SISO STS-SD with $L_{max} = 0$) can outperform LSD —in terms of ITCs— if $I_A$ is close to 1 and the list-size is small. We can therefore conclude that SISO STS-SD has a fundamental performance advantage over LSD, which is in agreement with the observations made in Section VI-C

## F. Approaching Outage-Capacity with SISO STS-SD

We finally compare the performance obtained with SISO STS-SD to outage capacity using the TGn type C channel model [38]. To this end, we define the $\varepsilon$-outage capacity $C_{out,\epsilon}$ as [46], [47]

$$P[I(\mathsf{SNR}, \mathcal{H}) < C_{out,\varepsilon}] = \varepsilon \qquad (46)$$

where $\mathcal{H} = \{\mathbf{H}[1], \ldots, \mathbf{H}[N]\}$ contains the $M_R \times M_T$ channel matrices for the $N = 64$ OFDM tones and [48]

$$I(\mathsf{SNR}, \mathcal{H}) = \frac{1}{N} \sum_{\ell=1}^{N} \log_2 \det \left( \mathbf{I}_{M_R} + \frac{\mathsf{SNR}}{M_T} \mathbf{H}^H[\ell] \mathbf{H}[\ell] \right).$$

The FER is lower-bounded by the outage probability (46) according to [49]

$$P[I(\mathsf{SNR}, \mathcal{H}) < RM_T Q] \leq \mathrm{FER}(\mathsf{SNR})$$

where the information rate per OFDM tone is given by $RM_T Q$. The performance comparison consists of setting the outage probability and the FER to 1% and identifying the corresponding SNR operating points. Fig. 12 shows the corresponding results for SISO STS-SD with different modulation schemes for $I = 1$ and $I = 8$. Note that the LLR clipping parameters are chosen so as to minimize complexity while retaining near-max-log optimal performance at 1% FER (i.e., we used $L_{\max} = 0.1$, $L_{\max} = 0.4$, $L_{\max} = 2.0$, and $L_{\max} = 6.0$ for 64-QAM, 16-QAM, QPSK, and BPSK, respectively). We can see that SISO STS-SD operates between 1.5 dB (for 4-QAM) and 5.3 dB SNR (for 64-QAM) away from outage capacity.

## VII. CONCLUSION

We proposed a soft-input soft-output MIMO detector based on single tree-search sphere decoding (STS-SD) as introduced in [7], [8]. Besides adapting the single-tree search paradigm to account for soft-inputs, key to obtaining low complexity of the proposed algorithm are tightening of the tree-pruning criterion, clipping of the extrinsic LLRs built into the tree search, and a novel method for incorporating compensation of self-interference in LLRs—caused by channel-matrix regularization—into the tree search. Finally, we proposed an LLR correction method, which was demonstrated to achieve substantial performance improvements at low additional computational complexity. Our simulation results showed that the SISO STS-SD algorithm offers a wide range of performance/complexity tradeoffs, clearly outperforms state-of-the-art SISO detectors for MIMO systems, and achieves close-to-optimal—in the sense of outage capacity—performance.

## REFERENCES

[1] B. M. Hochwald and S. ten Brink, "Achieving near-capacity on a multiple-antenna channel," *IEEE Trans. on Comm.*, vol. 51, no. 3, pp. 389–399, Mar. 2003.

[2] M. Tüchler, A. C. Singer, and R. Kötter, "Minimum mean squared error equalization using a priori information," *IEEE Trans. on Sig. Proc.*, vol. 50, no. 3, pp. 673–683, Mar. 2002.

[3] J. Boutros, N. Gresset, L. Brunel, and M. Fossorier, "Soft-input soft-output lattice sphere decoder for linear channels," in *Proc. of IEEE GLOBECOM*, vol. 3, San Francisco, CA, USA, Dec. 2003, pp. 1583–1587.

[4] B. Steingrimsson, T. Luo, and K. M. Wong, "Soft quasi-maximum-likelihood detection for multiple-antenna wireless channels," *IEEE Trans. on Sig. Proc.*, vol. 51, no. 11, pp. 2710–2719, Nov. 2003.

[5] S. Bäro, J. Hagenauer, and M. Witzke, "Iterative detection of MIMO transmission using a list-sequential (LISS) detector," in *Proc. of IEEE ICC*, vol. 4, Anchorage, AK, USA, May 2003, pp. 2653–2657.

[6] H. Vikalo, B. Hassibi, and T. Kailath, "Iterative decoding for MIMO channels via modified sphere decoder," *IEEE Trans. on Wireless Comm.*, vol. 3, no. 6, pp. 2299–2311, Nov. 2004.

[7] C. Studer, A. Burg, and H. Bölcskei, "Soft-output sphere decoding: Algorithms and VLSI implementation," *IEEE J-SAC*, vol. 26, no. 2, pp. 290–300, Feb. 2008.

[8] J. Jaldén and B. Ottersten, "Parallel implementation of a soft output sphere decoder," in *Proc. of Asilomar Conf. on Signals, Systems, and Computers*, Monterey, CA, USA, Nov. 2005, pp. 581–585.

[9] H. Vikalo and B. Hassibi, "Modified Fincke-Pohst algorithm for low-complexity iterative decoding over multiple antenna channels," in *Proc. of IEEE ISIT*, Lausanne, Switzerland, June 2002, p. 390.

[10] ——, "Towards closing the capacity gap on multiple antenna channels," in *Proc. of IEEE ICASSP*, Orlando, FL, USA, May 2002, pp. 2385–2388.

[11] A. Burg, "VLSI circuits for MIMO communication systems," Ph.D. dissertation, ETH Zürich, Switzerland, 2006.

[12] C. Studer, "Iterative MIMO decoding: Algorithms and VLSI implementation aspects," Ph.D. dissertation, ETH Zürich, Switzerland, 2009, (in preparation).

[13] U. Fincke and M. Pohst, "Improved methods for calculating vectors of short length in a lattice, including a complexity analysis," *Math. of Computation*, vol. 44, pp. 463–471, Apr. 1985.

[14] W. H. Mow, "Maximum likelihood sequence estimation from the lattice viewpoint," Master's thesis, Chinese University of Hong Kong, Department of Information Engineering, 1991.

[15] C. P. Schnorr and M. Euchner, "Lattice basis reduction: Improved practical algorithms and solving subset sum problems," *Math. Programming*, vol. 66, no. 2, pp. 181–191, Sept. 1994.

[16] E. Viterbo and E. Biglieri, "A universal decoding algorithm for lattice codes," in *14ème colloque GRETSI*, Juan-les-Pins, France, Sept. 1993, pp. 611–614.

[17] E. Viterbo and J. Boutros, "A universal lattice code decoder for fading channels," *IEEE Trans. on Inf. Th.*, vol. 45, no. 5, pp. 1639–1642, July 1999.

[18] O. Damen, A. Chkeif, and J.-C. Belfiore, "Lattice code decoder for space-time codes," *IEEE Comm. Letters*, vol. 4, no. 5, pp. 161–163, May 2000.

[19] R. Wang and G. B. Giannakis, "Approaching MIMO channel capacity with soft detection based on hard sphere decoding," *IEEE Trans. on Comm.*, vol. 54, no. 4, pp. 587–590, Apr. 2006.

[20] J. Jaldén and B. Ottersten, "On the complexity of sphere decoding in digital communications," *IEEE Trans. on Sig. Proc.*, vol. 53, no. 4, pp. 1474–1484, Apr. 2005.

[21] E. Agrell, T. Eriksson, A. Vardy, and K. Zeger, "Closest point search in lattices," *IEEE Trans. on Inf. Th.*, vol. 48, no. 8, pp. 2201–2214, Aug. 2002.

[22] A. Burg, M. Borgmann, M. Wenk, M. Zellweger, W. Fichtner, and H. Bölcskei, "VLSI implementation of MIMO detection using the sphere decoding algorithm," *IEEE JSSC*, vol. 40, no. 7, pp. 1566–1577, July 2005.

[23] M. Stojnic, H. Vikalo, and B. Hassibi, "Speeding up the sphere decoder with $H^\infty$ and SDP inspired lower bounds," *IEEE Trans. on Sig. Proc.*, vol. 56, no. 2, pp. 712–726, Feb. 2008.

[24] G. Caire, G. Taricco, and E. Biglieri, "Bit-interleaved coded modulation," *IEEE Trans. on Inf. Th.*, vol. 44, no. 3, pp. 927–946, May 1998.

[25] J. Hagenauer, E. Offer, and L. Papke, "Iterative decoding of binary block and convolutional codes," *IEEE Trans. on Inf. Th.*, vol. 42, no. 2, pp. 429–445, Mar. 1996.

[26] M. S. Yee, "Max-Log-Map sphere decoder," in *Proc. of IEEE ICASSP*, vol. 3, Philadelphia, PA, USA, Mar. 2005, pp. 1013–1016.

[27] D. Seethaler, H. Artés, and F. Hlawatsch, "Dynamic nulling-and-canceling for efficient near-ML decoding of MIMO systems," *IEEE Trans. on Sig. Proc.*, vol. 54, no. 12, pp. 4741–4752, Dec. 2006.

[28] S. W. Kim and K. P. Kim, "Log-likelihood-ratio-based detection ordering in V-BLAST," *IEEE Trans. on Comm.*, vol. 54, no. 2, pp. 302–307, Feb. 2006.

[29] D. Wübben, R. Böhnke, J. Rinas, V. Kühn, and K.-D. Kammeyer, "Efficient algorithm for decoding layered space-time codes," *IEE Electronics Letters*, vol. 37, no. 22, pp. 1348–1350, Oct. 2001.

[30] M. Damen, H. El Gamal, and G. Caire, "On maximum likelihood detection and the search for the closest lattice point," *IEEE Trans. on Inf. Th.*, vol. 49, no. 10, pp. 2389–2402, Oct. 2003.

[31] P. Luethi, A. Burg, S. Haene, D. Perels, N. Felber, and W. Fichtner, "VLSI implementation of a high-speed iterative sorted MMSE QR decomposition," in *Proc. of IEEE ISCAS*, New Orleans, LA, USA, May 2007, pp. 1421–1424.

[32] D. Wübben, R. Böhnke, V. Kühn, and K.-D. Kammeyer, "MMSE extension of V-BLAST based on sorted QR decomposition," in *Proc. of IEEE VTC-Fall*, vol. 1, Orlando, FL, USA, Oct. 2003, pp. 508–512.

[33] E. Zimmermann and G. Fettweis, "Unbiased MMSE tree search detection for multiple antenna systems," in *Proc. of WPMC*, San Diego, CA, USA, Sept. 2006.

[34] M. van Dijk, A. J. E. M. Janssen, and A. G. C. Koppelaar, "Correcting systematic mismatches in computed log-likelihood ratios," *European Trans. on Telecomm.*, vol. 14, no. 3, pp. 227–244, July 2003.

[35] A. Burg, M. Wenk, and W. Fichtner, "VLSI implementation of pipelined sphere decoding with early termination," in *Proc. of EUSIPCO*, Florence, Italy, Sept. 2006.

[36] C. Studer, D. Seethaler, and H. Bölcskei, "Finite lattice-size effects in MIMO detection," in *Proc. 42th Asilomar Conf. on Signals, Systems, and Computers*, Oct. 2008, pp. 2032–2037.

[37] A. Burg, M. Borgmann, M. Wenk, C. Studer, and H. Bölcskei, "Advanced receiver algorithms for MIMO wireless communications," in *Proc. of DATE*, vol. 1, Munich, Germany, Mar. 2006, pp. 593–598.

[38] V. Erceg *et al.*, *TGn channel models*, May 2004, IEEE 802.11 document 03/940r4.

[39] L. Bahl, J. Cocke, F. Jelinek, and J. Raviv, "Optimal decoding of linear codes for minimizing symbol error rate," *IEEE Trans. on Inf. Th.*, vol. 20, no. 2, pp. 284–287, Mar. 1974.

[40] M. Wenk, A. Burg, M. Zellweger, C. Studer, and W. Fichtner, "VLSI implementation of the list sphere algorithm," in *Proc. of 24th NORCHIP Conf.*, Linköping, Sweden, Nov. 2006, pp. 107–110.

[41] J. Hagenauer and C. Kuhn, "The list-sequential (LISS) algorithm and its application," *IEEE Trans. on Comm.*, vol. 55, no. 5, pp. 918–928, May 2007.

[42] *3rd Generation Partnership Project; Technical Specification Group Radio Access Network; Multiplexing and Channel Coding (FDD)*, 3GPP Organizational Partners TS 25.212, Rev. 5.10.0, June 2005.

[43] J. Boutros, F. Boixadera, and C. Lamy, "Bit-interleaved coded modulations for multiple-input multiple-output channels," in *Proc. of IEEE ISSSTA*, vol. 1, Sept. 2000, pp. 123–126.

[44] S. ten Brink, "Convergence behavior of iteratively decoded parallel concatenated codes," *IEEE Trans. on Comm.*, vol. 49, pp. 1727–1737, Oct. 2001.

[45] H. Bölcskei, D. Gesbert, C. Papadias, and A. J. van der Veen, Eds., *Space-Time Wireless Systems: From Array Processing to MIMO Communications*. Cambridge Univ. Press, 2006.

[46] İ. E. Telatar, "Capacity of multi-antenna Gaussian channels," *European Trans. on Telecomm.*, vol. 10, no. 6, pp. 585–596, 1999.

[47] E. Biglieri, J. Proakis, and S. Shamai, "Fading channels: Information-theoretic and communications aspects," *IEEE Trans. on Inf. Th.*, vol. 44, no. 6, pp. 2619–2692, Oct. 1998.

[48] H. Bölcskei, D. Gesbert, and A. J. Paulraj, "On the capacity of OFDM-based spatial multiplexing systems," *IEEE Trans. on Comm.*, vol. 50, no. 2, pp. 225–234, Feb. 2002.

[49] L. Zheng and D. N. C. Tse, "Diversity and multiplexing: A fundamental tradeoff in multiple-antenna channels," *IEEE Trans. on Inf. Th.*, vol. 5, no. 49, pp. 1073–1096, May 2003.

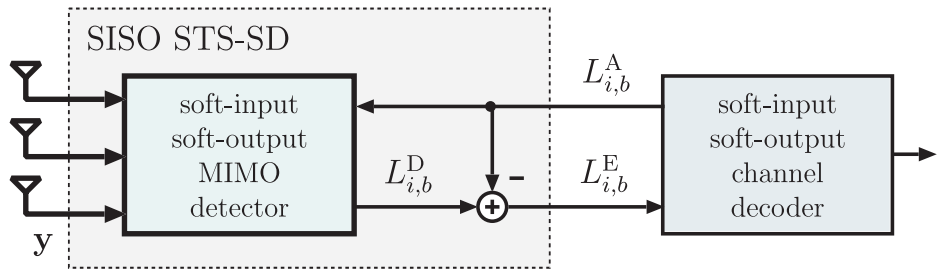

Fig. 1. Iterative MIMO decoder. SISO STS-SD (corresponding to the dashed box) directly computes extrinsic LLRs.

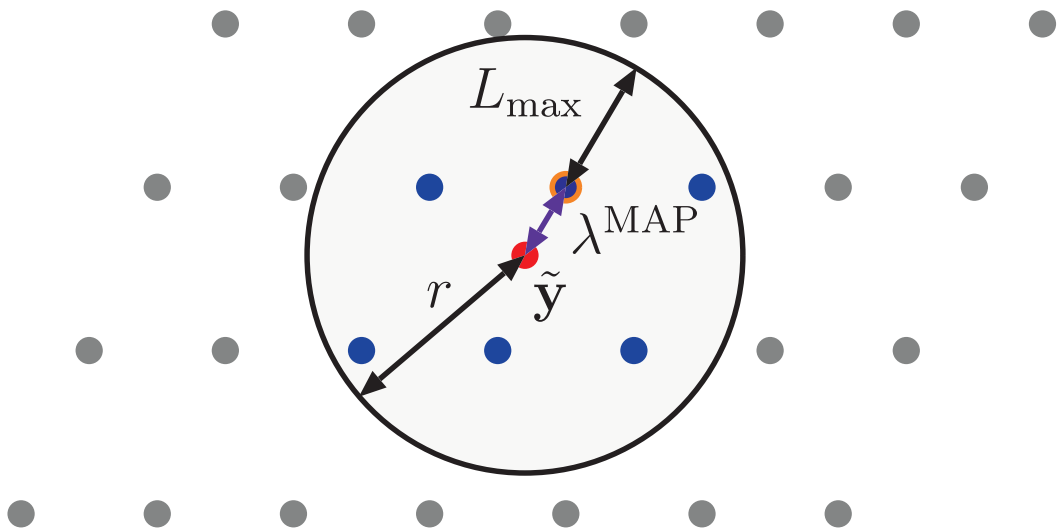

Fig. 2.   Illustration of extrinsic LLR clipping incorporated into the tree search. The constraint $|L^{\mathrm{E}}_{i,b}| \leq L_{\max}$ translates into a radius constraint $r = \sqrt{\lambda^{\mathrm{MAP}} + L_{\max}}$ for the search sphere.

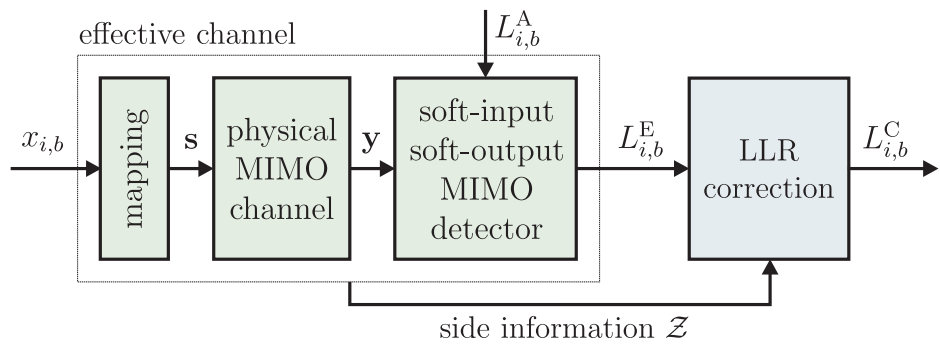

Fig. 3. LLR correction post-processes the LLRs resulting from the effective channel using side information $\mathcal{Z}$.

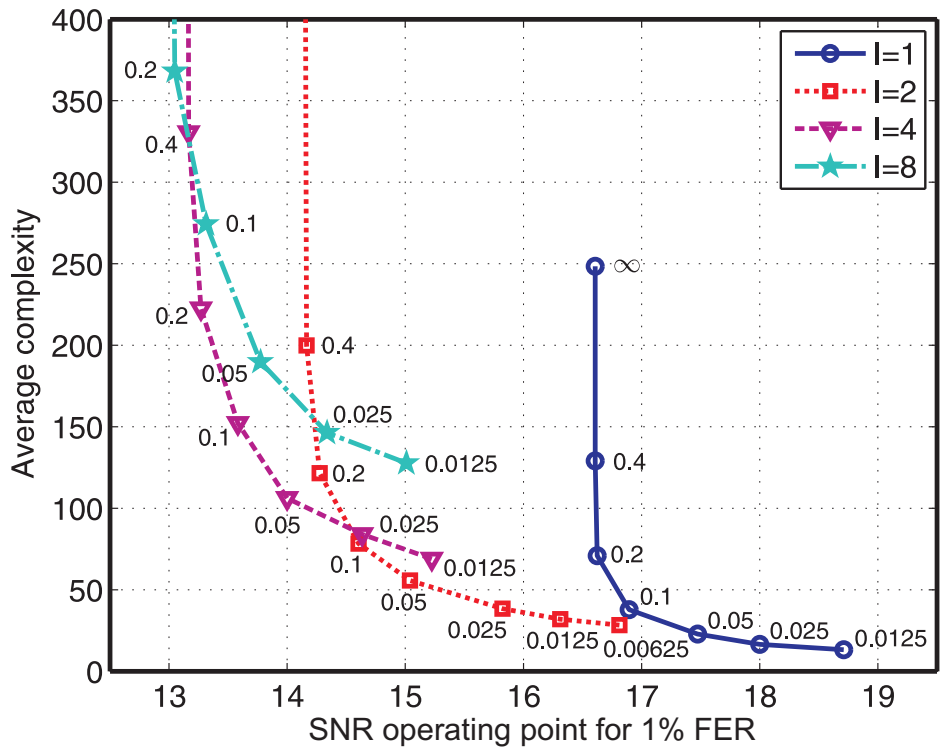

Fig. 4. Performance/complexity tradeoff of SISO STS-SD with SQRD. The numbers next to the curves correspond to normalized LLR clipping parameters.

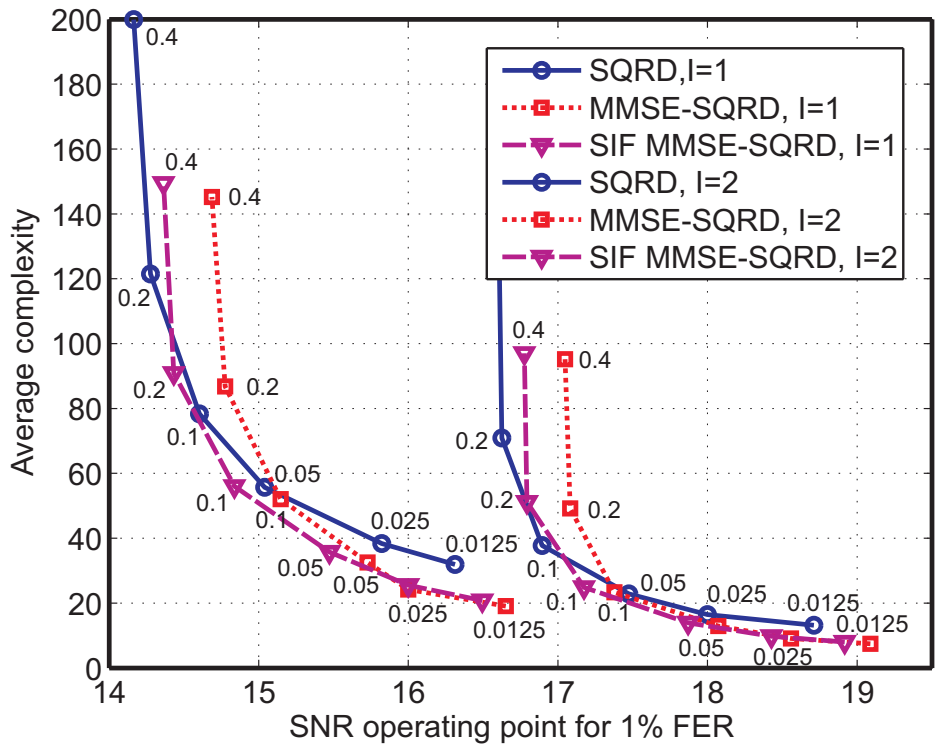

Fig. 5.   Performance/complexity tradeoff of SISO STS-SD with SQRD, MMSE-SQRD, and SIF MMSE-SQRD. The numbers next to the curves correspond to normalized LLR clipping parameters.

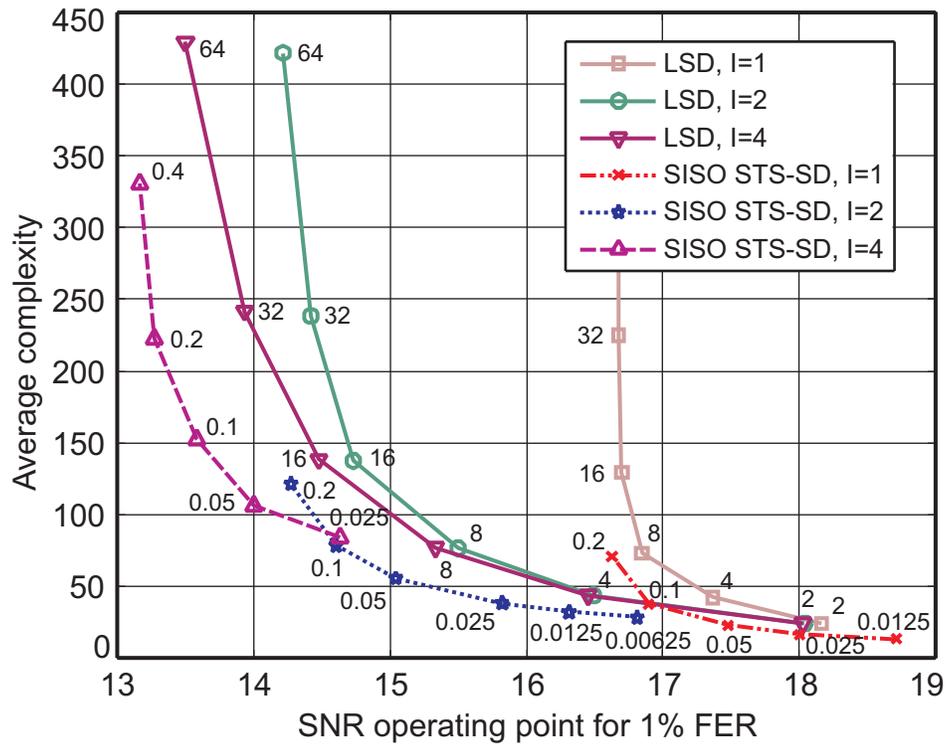

Fig. 6. Performance/complexity tradeoff of LSD [1] and SISO STS-SD, both using SQRD. The numbers next to the curves correspond to the list size for LSD and to normalized LLR clipping parameters for SISO STS-SD.

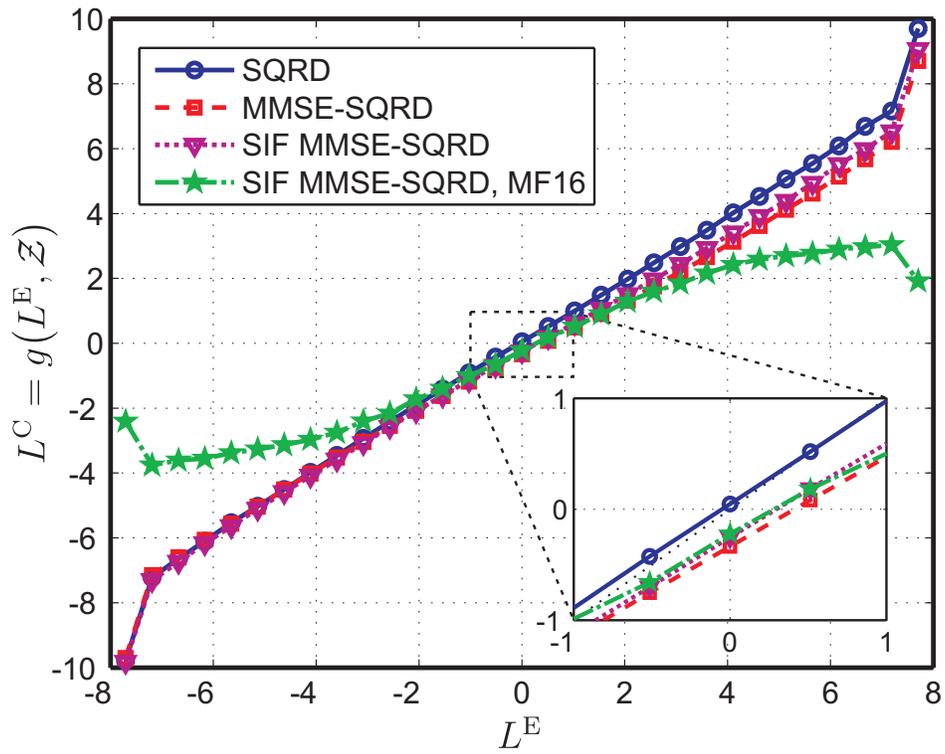

Fig. 7. Different LLR correction functions for $I = 1$ at $\mathsf{SNR} = 16\,\mathrm{dB}$ using $\mathcal{Z} = \{L_{\max}, D_{\mathrm{avg}}, \mathsf{SNR}, T\}$.

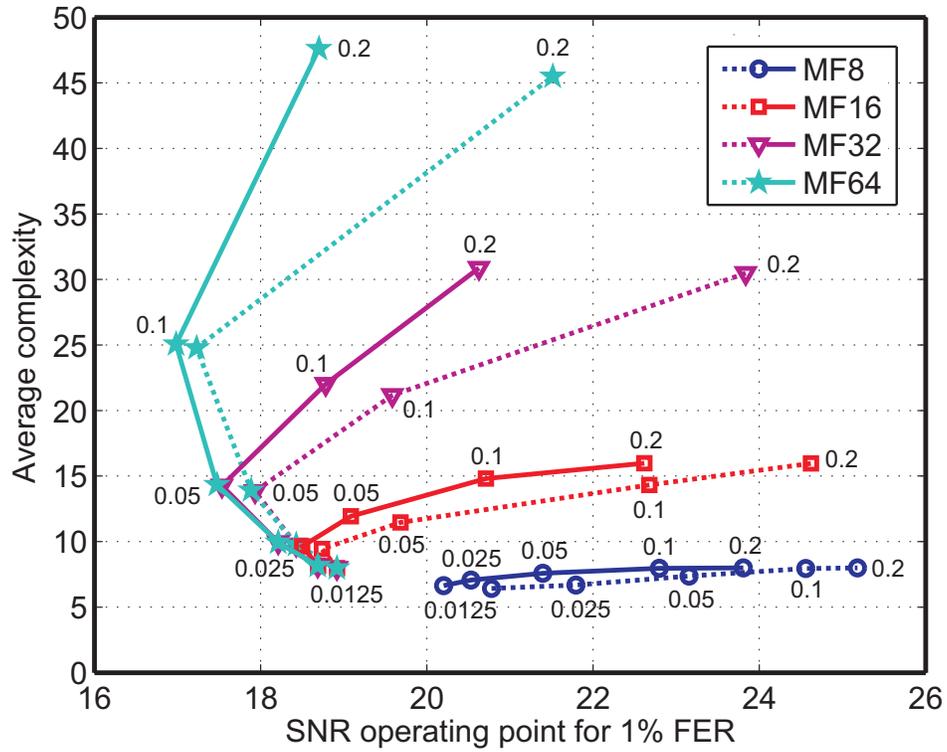

(a) $I = 1$

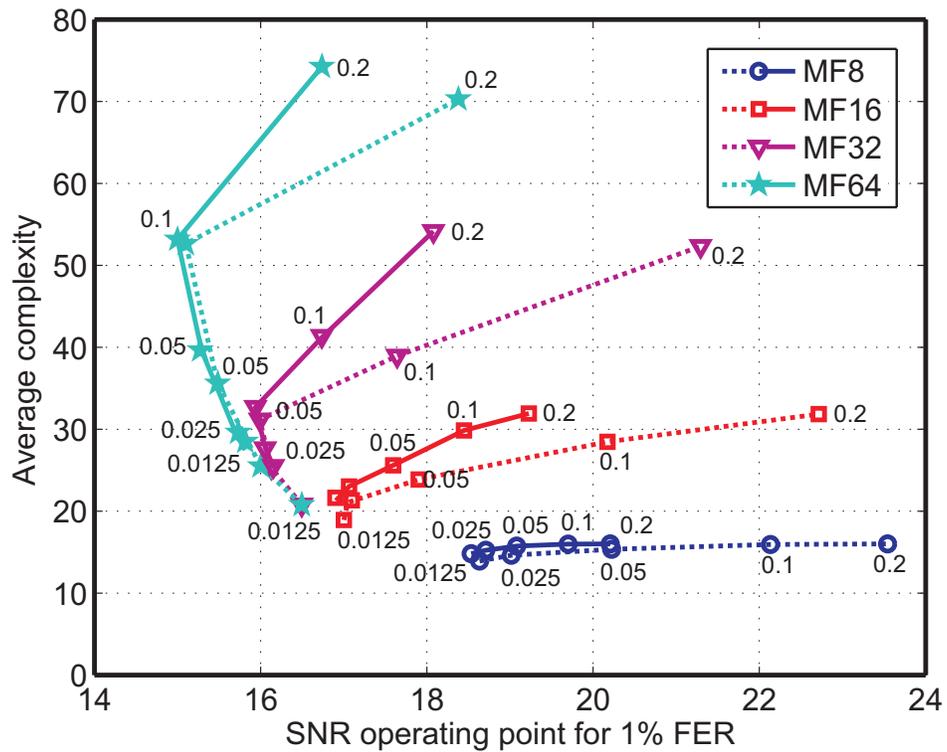

(b) $I = 2$

Fig. 8. Impact of LLR correction. The solid lines correspond to the performance obtained with LLR correction, whereas the dotted lines pertain to un-corrected LLRs. Both variants employ early termination with MF scheduling and compensation of self-interference in the LLRs in combination with MMSE-SQRD.

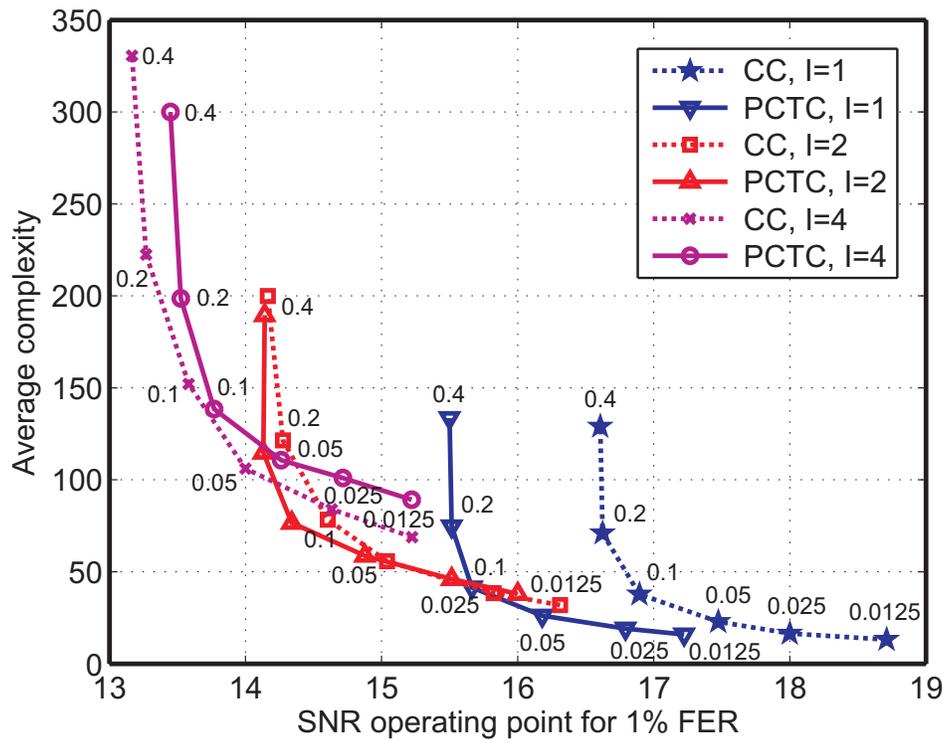

Fig. 9.   Performance/complexity tradeoff of SISO STS-SD with SQRD (without regularization). Comparison between parallel-concatenated turbo codes (PCTCs) and convolutional codes (CCs).

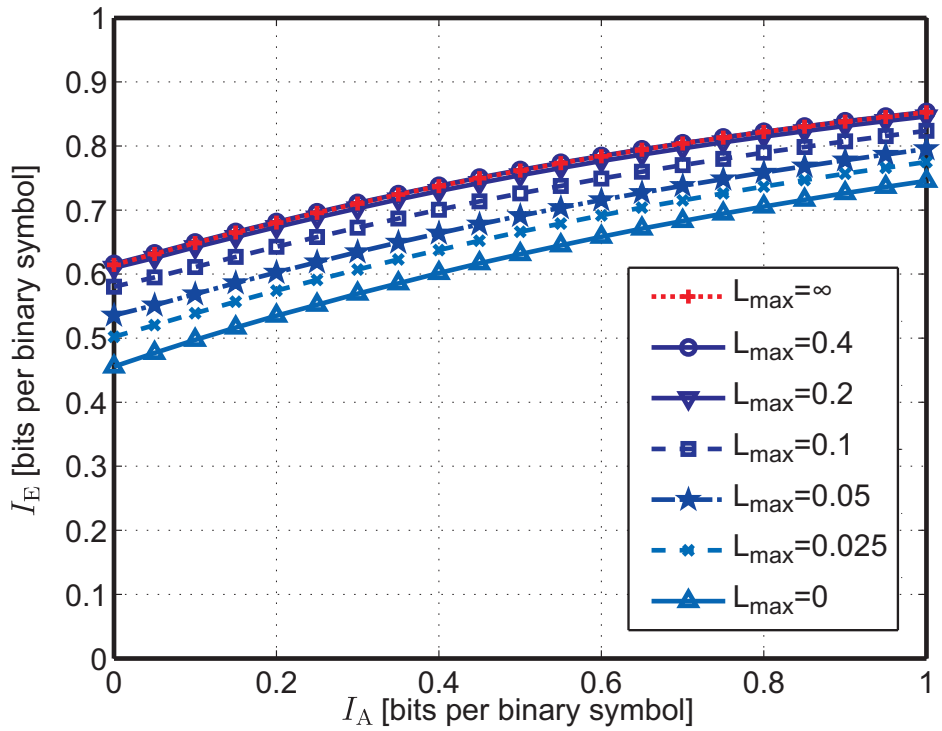

Fig. 10.   ITC of SISO STS-SD at SNR = 12 dB for different (normalized) LLR clipping parameters.

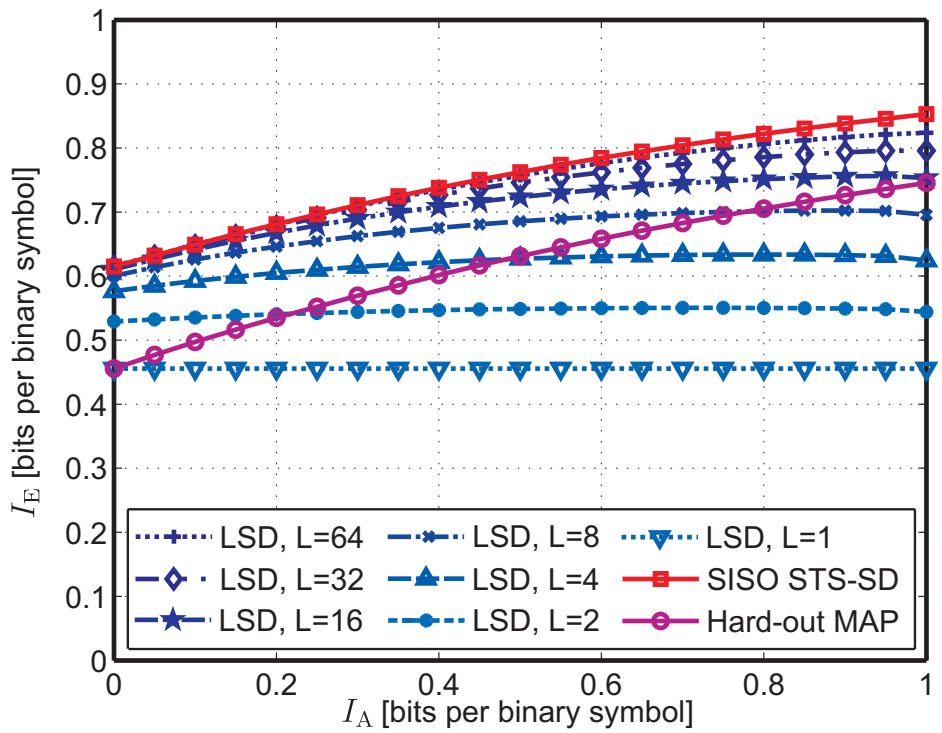

Fig. 11. ITC of LSD compared to hard-output MAP and (max-log) optimal SISO STS-SD performance at SNR = 12 dB.

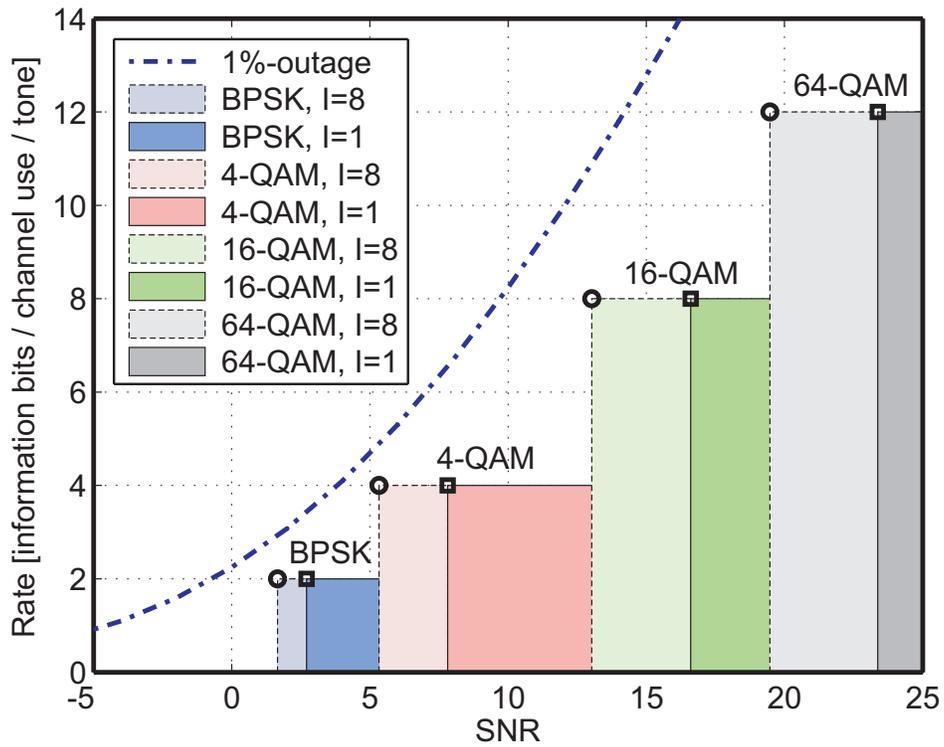

Fig. 12.  1%-outage-capacity compared to the SNR operating points of SISO STS-SD for 1% FER. Squares and circles correspond to SNR operating points realized by $I = 1$ and $I = 8$, respectively.

TABLE I

AVERAGE COMPLEXITY REDUCTION OBTAINED BY TIGHTENING OF THE TREE-PRUNING CRITERION BASED ON THE
EUCLIDEAN-DISTANCE TERM ONLY

| SNR | $L_{\max}$ | std. [nodes] | tight [nodes] | reduction |
|------|------|------|------|------|
| 10 dB | 0.0125 | 34.9 | 34.4 | 1.4% |
|        | $\infty$ | 328.3 | 327.8 | 0.2% |
| 20 dB | 0.0125 | 11.0 | 10.8 | 1.8% |
|        | $\infty$ | 227.2 | 227.0 | 0.1% |

TABLE II

Average complexity reduction obtained by tightening of the tree-pruning criterion based on the prior term only

| SNR | $I$ | $L_{\max}$ | std. [nodes] | tight [nodes] | reduction |
|---|---|---|---|---|---|
| 10 dB | 1 | 0.0125 | 1890.4 | 34.9 | 98.2% |
|  |  | $\infty$ | 2440.2 | 328.3 | 86.5% |
|  | 2 | 0.0125 | 1630.6 | 43.4 | 97.3% |
|  |  | $\infty$ | 2148.4 | 406.6 | 81.1% |
| 20 dB | 1 | 0.0125 | 1914.7 | 11.0 | 99.4% |
|  |  | $\infty$ | 2397.0 | 227.2 | 90.5% |
|  | 2 | 0.0125 | 1228.7 | 6.2 | 99.5% |
|  |  | $\infty$ | 361.9 | 132.4 | 65.9% |